\documentclass[aps,pre,reprint,superscriptaddress,longbibliography,amsmath,amssymb]{revtex4-2}

\usepackage{amsmath,amssymb}
\usepackage{bm}
\usepackage{graphicx}
\usepackage{booktabs}
\usepackage{upgreek}
\usepackage{xcolor}
\usepackage[colorlinks=true,allcolors=blue]{hyperref}
\usepackage[nameinlink]{cleveref}
\crefname{equation}{Eq.}{Eqs.}
\Crefname{equation}{Equation}{Equations}
\creflabelformat{equation}{#2(#1)#3}
\crefrangelabelformat{equation}{#3(#1)#4--#5(#2)#6}
\usepackage{orcidlink}

\graphicspath{{figures/}}

\newcommand{\jj}{\mathrm{j}}
\newcommand{\Real}{\operatorname{Re}}
\newcommand{\Imag}{\operatorname{Im}}
\newcommand{\PV}{\operatorname{P.V.}}
\newcommand{\dd}{\mathrm{d}}
\newcommand{\bX}{\mathbf{X}}
\newcommand{\bk}{\mathbf{k}}
\newcommand{\bp}{\mathbf{p}}
\newcommand{\bs}{\mathbf{s}}
\newcommand{\bu}{\mathbf{u}}
\newcommand{\bv}{\mathbf{v}}
\newcommand{\bG}{\mathbf{G}}
\newcommand{\bM}{\mathbf{M}}
\newcommand{\aC}{a_\mathrm{C}}
\newcommand{\vT}{V_\mathrm{0T}}
\newcommand{\vL}{V_\mathrm{0L}}
\newcommand{\kT}{k_\mathrm{0T}}
\newcommand{\kL}{k_\mathrm{0L}}
\newcommand{\us}{\upmu\text{s}}

\begin{document}
\setlength{\emergencystretch}{1.5em}

\title{Multiplicity of elastic wave propagation in random polycrystals}

\author{Ming Huang\,\orcidlink{0000-0002-5868-4244}}
\email{ming.huang@sussex.ac.uk}
\affiliation{Department of Engineering and Design, University of Sussex, Brighton BN1 9RH, United Kingdom}
\affiliation{Department of Mechanical Engineering, Imperial College London, London SW7 2AZ, United Kingdom}
\author{Stanislav I. Rokhlin\,\orcidlink{0009-0002-5791-1430}}
\affiliation{Department of Materials Science and Engineering, The Ohio State University, 1248 Arthur E. Adams Drive, Columbus, Ohio 43221, USA}
\author{Michael J. S. Lowe}
\affiliation{Department of Mechanical Engineering, Imperial College London, London SW7 2AZ, United Kingdom}

\begin{abstract}
The coherent elastic wave in untextured polycrystals, central to ultrasonic and seismic characterization, was recently observed to have two modes per polarization at short wavelengths. However, their physical origin and amplitude partition have remained open. We address both by theory, grain-scale numerical simulations and a grain velocity estimate. The first-order smoothing approximation, simplified by the far-field approximation, predicts three waves, which are the classical effective wave and one for each type of scattered wave (longitudinal and transverse). At long wavelengths, only the effective wave is weakly attenuated. At short wavelengths, one of the other two waves joins it with comparable attenuation, whereas the third stays far more attenuated. A new residue method, which needs no fitting, gives the same pattern for the amplitude partition, with the two waves sharing the coherent field and the third carrying little. The simulations reveal exactly these two modes, for transverse waves in iron and longitudinal waves in strongly anisotropic SrMgSn, and never the third. The wavefields reveal the mechanism, in which the incident wave is extinguished within one grain correlation length and the coherent field is rebuilt by forward multiple scattering. Strikingly, a grain velocity estimate from single-crystal constants provides independent evidence, tracing the two modes to plane waves of the individual grains and closely reproducing their velocities. Observability of multiple waves is thus set by attenuation and amplitude partition together, and short-wavelength measurements in random media must allow for more than one coherent mode.
\end{abstract}


\maketitle

\section{Introduction}\label{sec:intro}

Elastic waves traveling through polycrystalline metals, rocks and ceramics are scattered by the grains, whose crystallographic orientations vary at random. The scattering attenuates the incident wave, by spreading its energy into an incoherent field, and changes its phase velocity. These effects underlie a wide range of practical measurements. In nondestructive evaluation, the attenuation and velocity of ultrasound are used to infer grain size and texture and to decide how small a defect can be detected in safety-critical components \citep{Papadakis1965,Thompson2008,Lan2015b,Lan2015a,Rokhlin2021}. In seismology, the same physics governs the propagation of high-frequency waves through the heterogeneous crust and mantle \citep{Sato2012}, and the scattering of seismic waves by the anisotropic iron grains of the Earth's inner core is used to constrain their size and texture \citep{Calvet2008,Monnereau2010,Calvet2016,Pang2023}. In most cases, the measured quantity is the coherent wave, i.e.\ the part of the field that survives averaging over the randomness of the microstructure, and its interpretation rests on a description of how this wave propagates.

The standard description is that of an effective medium, in which the coherent wave is characterized by a single complex effective wavenumber for each polarization as in other random media \citep{Ishimaru1978,Sheng2006,Akkermans2007}. Its real part gives the phase velocity, and its imaginary part gives the scattering-induced attenuation. This picture has been developed for polycrystals over several decades \citep{Papadakis1965,Hirsekorn1982,Stanke1984,Weaver1990,Turner1999,Calvet2012,Rokhlin2015,Kube2015,Sha2020}. Its modern form rests on the Dyson equation for the mean Green's function, with the mass operator approximated to second order in the fluctuations of the elastic constants by the first-order smoothing approximation \citep[FOSA;][]{Stanke1984,Weaver1990}. Three treatments follow from it: the Born approximation, which evaluates the mass operator at the unperturbed wavenumber \citep{Weaver1990}; the far-field approximation (FFA), which simplifies the mass operator \citep{Rokhlin2015}; and the spectral function (the imaginary part of the mean Green's function at real wavenumbers), in which the coherent waves appear as peaks \citep{Calvet2012}. The picture has been validated by experiment and, more recently, by grain-scale numerical simulation \citep{VanPamel2015,VanPamel2017,Ryzy2018,Huang2020a,Huang2022a,Huang2022b}. Three frequency regimes are usually distinguished: the Rayleigh regime, where the wavelength is much longer than the grain size; the stochastic regime, where the two are comparable; and the geometric regime at higher frequencies, where waves cross the grains much like rays. These measurements and simulations cover the first two regimes and have revealed only one coherent wave per polarization. To our knowledge, the geometric regime had not been reached in laboratory or grain-scale numerical experiments before our recent work \citep{Huang2026prl}.

However, other random media are known to carry more than one coherent wave of a given polarization. Biot predicted this first for fluid-saturated porous solids, whose coupled solid--fluid motion supports fast and slow longitudinal waves \citep{Biot1956a,Biot1956b}, and Plona observed the slow wave in ultrasonic experiments more than two decades later \citep{Plona1980}. Fast and slow longitudinal waves are now routinely resolved in cancellous bone \citep{Hoffman2012,Meziere2016}. For colloidal suspensions of spheres, Brillouin-scattering measurements revealed two longitudinal modes once the wavelength became comparable to the sphere size \citep{Liu1990}. These were explained as quasimodes (peaks of the spectral function with finite lifetime) associated with sphere antiresonances and coupled interfacial waves \citep{Jing1991,Jing1992} and were later mapped in detail by pulsed ultrasonics \citep{Page1996,Cowan2011}. Most recently, it has been proved for random particulate composites that the ensemble-averaged field is a superposition of infinitely many effective waves rather than one wave \citep{Gower2019a,Gower2019b,Gower2021}, and an independent variational treatment found that a layer of random composite supports exactly two mean plane waves \citep{Willis2024}. The multiplicity in all of these systems is nevertheless tied to the presence of two constituents, such as a solid frame saturated by a pore fluid or discrete scatterers embedded in a host.

Single-phase polycrystals pose the question in its purest form. There is no second constituent, and the only disorder is the random orientation of elastically anisotropic grains. Theory has long suggested that such media also carry more than one wave. For scalar waves in continuous random media with an exponential correlation function, \citet{Frisch1968} follows \citet{Tatarskii1963} to give the mean Green's function of the FOSA as the sum of two terms that each have their own complex wavenumber and amplitude \citep[see also][]{Rytov1989}. A polycrystal is an elastic example of such media. \citet{Calvet2012} showed that a second wave is required by causality because a single wave with the asymptotic velocity and attenuation of the geometric limit cannot satisfy the Kramers--Kronig relations, which link velocity dispersion and attenuation in any causal medium \citep{WeaverPao1981,ODonnell1981,Beltzer1986}. For elastic waves, the dispersion equations of the FOSA \citep{Stanke1984} and of the FFA were found to have two longitudinal and two transverse roots \citep{Rokhlin2015}, and a third root of each polarization was later found in the same equations \citep{Huang2021}. Each root is the complex wavenumber of a possible coherent wave. Throughout this paper, roots refer to such solutions of a theory, and modes refer to the coherent waves that are observed in experiments or simulations. Despite these predictions, this multiplicity long remained purely theoretical in polycrystals. No additional mode had been observed, and it was unknown whether the extra roots are physical waves or artifacts of the approximations that produce them.

Our recent work \citep{Huang2026prl} answered this question of existence. By examining the coherent field in the frequency--wavenumber plane, grain-scale finite-element simulations of three-dimensional polycrystals resolved two coherent modes for both transverse and longitudinal waves once the wavelength became comparable to the grain correlation length. The measured modes follow the roots of the dispersion equation, which always carries a third root of larger attenuation that is never observed. The extra roots are therefore physical, and the number of observable modes was attributed there to the attenuation hierarchy of the roots.

Two further questions remain to be addressed formally. The first is the physical origin of the modes, i.e.\ why the dispersion equation has three roots, what wave each root represents and what the two observable modes are. The causality argument of \citet{Calvet2012} shows why a second root is needed, but not which root it is or what it represents. The question is sharpened by the fact that the theory is a low-order approximation that is formally valid for weak scattering, whereas the second mode appears where the scattering is strong. An account of the modes that does not rest on the scattering theory is therefore needed as well. The second question is how the coherent field is partitioned among the roots. Solving the dispersion equation of the FOSA or of the FFA gives the velocity and attenuation that each root predicts but not its amplitude. The spectral function does contain amplitude information, but it has been applied to polycrystals only by fitting Lorentzian peaks, and a root that is too damped to form a peak of its own cannot be resolved in this way. Without the amplitudes, it remains unclear whether attenuation alone decides which of the roots are observed.

This paper addresses both questions by theory, grain-scale finite-element simulations and a grain velocity estimate, and it makes five contributions. First, by writing the FOSA mass operator as a sum over scattering terms and simplifying it with the FFA, we show that the dispersion equation has exactly three roots, trace each root to the incident wave or to one scattering term, and identify what the two observable modes represent physically. Second, we introduce a residue method that obtains the amplitude partition of every root from the spectral function without fitting, including a root too damped to form a peak of its own. Third, the simulated wavefields reveal the mechanism behind the multiple modes, in which the incident plane wave is extinguished and the coherent field is rebuilt by forward multiple scattering, and the modes are measured in the frequency--wavenumber plane. Fourth, a grain velocity estimate based only on the plane waves of the single crystal reproduces the velocities of the observed modes and thus confirms them independently of the scattering theory. Fifth, the comparison of theory and simulation establishes which roots are observed and shows that attenuation and amplitude partition together decide observability. Beyond polycrystals, the results apply to any random medium described by the Dyson equation and bear directly on ultrasonic characterization and seismic imaging at short wavelengths, where measured velocities and attenuations must be interpreted in terms of more than one coherent mode.

The paper is organized as follows. Section~\ref{sec:theory} develops the theory from the FOSA dispersion equation through the three roots of the FFA to the amplitude partition. Section~\ref{sec:simulation} describes the simulations and the extraction of the modes. Section~\ref{sec:ray} presents the grain velocity estimate of the modes in the geometric regime. Section~\ref{sec:results} compares theory, simulation and the grain velocity estimate for transverse waves in iron and longitudinal waves in SrMgSn and discusses the results, and Section~\ref{sec:conclusions} concludes.

\section{Theory}\label{sec:theory}

This section develops the theory. Section~\ref{sec:fosa} derives the dispersion equation of the FOSA \citep{Stanke1984,Weaver1990} and defines its scattering channels, Section~\ref{sec:ffa} uses the FFA to show that it has three roots and to identify what each represents, and Section~\ref{sec:spectral} obtains the amplitude of every root from the spectral function.

\subsection{Mean Green's function and the first-order smoothing approximation}
\label{sec:fosa}

We consider a polycrystal formed by densely packed, fully bonded single-phase grains. The mass density $\rho$ is constant throughout, and the medium is described by the elastic tensor $c_{ijkl}(\bX)$, which varies with position $\bX$ because the crystallographic orientation changes from grain to grain. Fields vary in time as $\exp(-\jj\omega t)$, with $\omega=2\pi f$ the angular frequency and $\jj^2=-1$. Wave propagation is governed by the equation of motion for the Green's function $G_{k\alpha}(\bX,\bX';\omega)$, which is proportional to the displacement in direction $k$ at $\bX$ due to a point force in direction $\alpha$ at $\bX'$ \citep{Karal1964,Weaver1990,Rokhlin2015},
\begin{equation}\label{eq:motion}
\begin{split}
 &\left[\frac{\partial}{\partial X_i}\,\frac{c_{ijkl}(\bX)}{\rho}\,\frac{\partial}{\partial X_l}
 +\omega^2\delta_{jk}\right]G_{k\alpha}(\bX,\bX';\omega)\\
 &\qquad=\delta_{j\alpha}\,\delta^3(\bX-\bX'),
\end{split}
\end{equation}
where $\delta_{j\alpha}$ is the Kronecker delta and $\delta^3$ the three-dimensional delta function. The elastic tensor is split into the Voigt average $C_{ijkl}=\langle c_{ijkl}\rangle$, which defines a homogeneous reference medium, and a zero-mean fluctuation,
\begin{equation}\label{eq:split}
 c_{ijkl}(\bX)=C_{ijkl}+\delta c_{ijkl}(\bX),\qquad \langle\delta c_{ijkl}\rangle=0,
\end{equation}
where $\langle\cdot\rangle$ denotes the ensemble average over grain orientations. For the untextured polycrystals considered here, orientations are uniformly distributed, the reference medium is isotropic, and its longitudinal and transverse velocities are the Voigt velocities $\vL$ and $\vT$. Other reference media can be used, as in the self-consistent approach of \citet{Kube2015sc}, but we take the Voigt average because the resulting theory agrees well with finite-element simulations \citep{Huang2022a,Huang2022b}.

\Cref{eq:motion} has random coefficients and cannot be solved in general. For the mean Green's function $\langle G_{k\alpha}\rangle$, it is replaced by the Dyson integral equation \citep{Frisch1968,Weaver1990},
\begin{equation}\label{eq:dyson}
\begin{split}
 \langle G_{k\alpha}(\bX,\bX')\rangle={}&G^0_{k\alpha}(\bX,\bX')
 +\iint G^0_{k\beta}(\bX,\mathbf{Y})\\
 &\times M_{\beta j}(\mathbf{Y},\mathbf{Z})\,\langle G_{j\alpha}(\mathbf{Z},\bX')\rangle\,\dd^3Y\,\dd^3Z,
\end{split}
\end{equation}
where $G^0$ is the bare Green's function of the reference medium (the solution of \cref{eq:motion} with $c_{ijkl}$ replaced by $C_{ijkl}$) and $M_{\beta j}$ is the mass operator, which contains all multiple-scattering events.

\Cref{eq:dyson} is exact, but the difficulty is that the mass operator is unknown. We apply the first-order smoothing approximation (FOSA), which is also called the Bourret approximation \citep{Bourret1962,Frisch1968,Rytov1989}, to keep only the leading term of its expansion in powers of the fluctuations, $\bM\simeq\langle\delta\mathcal{L}\,\bG^0\,\delta\mathcal{L}\rangle$, where $(\delta\mathcal{L})_{jk}=\partial_i\,(\delta c_{ijkl}/\rho)\,\partial_l$ is the fluctuating part of the operator in \cref{eq:motion}. Each factor $\delta\mathcal{L}$ is a scattering event and $\bG^0$ carries the wave between two events. The approximation is second order in the fluctuations, but the mean Green's function obtained through \cref{eq:dyson} contains every chain of such pairs of events to all orders. It neglects the terms in which three or more events are correlated, such as recurrent scattering, in which the wave returns to a grain it has already visited \citep{Roy2025}. The FOSA is formally valid for weak scattering, which means that the coherent wave is attenuated little over one correlation length of the grains. However, \citet{Calvet2012} and \citet{Rokhlin2015} have shown that it captures the transition to the geometric regime and remains qualitatively correct, as our comparison with numerical simulation confirms in this work.

The only statistical property of the medium that enters the FOSA mass operator is the average of the product of the fluctuations at two points, $\langle\delta c_{ijkl}(\bX)\,\delta c_{\alpha\beta\gamma\delta}(\bX')\rangle$. Two assumptions reduce it to a simple form \citep{Stanke1984,Weaver1990}. First, the orientations of different grains are uncorrelated and independent of the grain shapes, so the average factorizes into a tensorial part and a spatial part. Second, the statistics are homogeneous, so that the spatial part depends only on the separation $\mathbf{r}=\bX-\bX'$. With these two assumptions, we have
\begin{equation}\label{eq:covariance}
 \langle\delta c_{ijkl}(\bX)\,\delta c_{\alpha\beta\gamma\delta}(\bX')\rangle
 =\langle\delta c_{ijkl}\,\delta c_{\alpha\beta\gamma\delta}\rangle\,w(\mathbf{r}).
\end{equation}
Here $\langle\delta c_{ijkl}\,\delta c_{\alpha\beta\gamma\delta}\rangle=\langle c_{ijkl}c_{\alpha\beta\gamma\delta}\rangle-C_{ijkl}C_{\alpha\beta\gamma\delta}$ is the eighth-rank covariance of the single-crystal constants, evaluated at one point, and $w(\mathbf{r})$ is the geometric two-point correlation function that gives the probability of two points separated by $\mathbf{r}$ lying in the same grain. For equiaxed grains, $w$ depends on $r=|\mathbf{r}|$ only, and we take the classical single exponential $w(r)=\exp(-r/a)$, where the correlation radius $a$ is also the correlation length $\int_0^\infty w\,\dd r$.

Because the statistics are homogeneous, \cref{eq:dyson} becomes algebraic in the spatial Fourier domain,
\begin{equation}\label{eq:dyson_k}
 \langle\bG(\bk)\rangle=\left[\bG^0(\bk)^{-1}-\bM(\bk)\right]^{-1},
\end{equation}
with $\bk=k\bp$ the wave vector and $\bp$ the unit propagation direction. After placing the covariance of \cref{eq:covariance} between the two gradient operators, the FOSA mass operator becomes a convolution over the scattered wave vector $\bk^S=k^S\bs$ \citep{Weaver1990,Rokhlin2015},
\begin{equation}\label{eq:massop}
\begin{split}
 M_{\beta m}(\bk)={}&k_\alpha k_l\int G^0_{\gamma j}(\bk^S)\,k^S_\delta k^S_i\\
 &\times\frac{\langle\delta c_{ijml}\,\delta c_{\alpha\beta\gamma\delta}\rangle}{\rho^2}\,
 W(\bk-\bk^S)\,\dd^3k^S,
\end{split}
\end{equation}
where the two-point correlation function enters through its spectral density $W$, which for the single exponential is
\begin{equation}\label{eq:W}
 W(\mathbf{q})=\frac{1}{(2\pi)^3}\int w(\mathbf{r})\exp(-\jj\mathbf{q}\cdot\mathbf{r})\,\dd^3r
 =\frac{a^3}{\pi^2\left(1+a^2q^2\right)^2}.
\end{equation}

The bare Green's function of the isotropic reference medium decomposes into three propagating waves $M=\mathrm{L},\mathrm{T1},\mathrm{T2}$ with unit polarization vectors $\bu^M$ and velocities $V_{0M}$ obtained from the Christoffel equation of the reference medium for the direction $\bp$ \citep{Turner1999,Rokhlin2015},
\begin{equation}\label{eq:G0}
 \bG^0(\bk)=\sum_M\frac{\bu^M\bu^M}{\omega^2-k^2V_{0M}^2}=\sum_M g^0_M\,\bu^M\bu^M,
\end{equation}
where $V_{0\mathrm{L}}=\vL$ and $V_{0\mathrm{T1}}=V_{0\mathrm{T2}}=\vT$ for the isotropic reference medium. The mean Green's function and the mass operator admit the same dyadic expansion, $\langle\bG\rangle=\sum_M g_M\bu^M\bu^M$ and $\bM=\sum_M m_M\bu^M\bu^M$, and thus \cref{eq:dyson_k} gives for each polarization
\begin{equation}\label{eq:gM}
 g_M(k,\omega)=\frac{1}{\omega^2-k^2V_{0M}^2-m_M(k,\omega)}\equiv\frac{1}{D_M(k,\omega)}.
\end{equation}
The dispersion equation for the complex effective wavenumber of a propagating wave $M$ is the vanishing of the denominator,
\begin{equation}\label{eq:dispersion}
 D_M(k,\omega)=\omega^2-k^2V_{0M}^2-m_M(k,\omega)=0.
\end{equation}
Each root $k=\Real k+\jj\Imag k$ with positive real and imaginary parts represents a forward-propagating coherent wave with phase velocity $V_M=\omega/\Real k$ and attenuation coefficient $\alpha_M=\Imag k$, which we refer to as the velocity and attenuation of the root. For equiaxed cubic grains, \cref{eq:dispersion} is the dispersion equation of \citet{Stanke1984} in a different form, as \citet{Rokhlin2015} verified numerically, and it is equivalent to that of \citet{Calvet2012}. For scalar waves, it reduces to the equations of \citet{Bourret1962} and \citet{Karal1964}.

The components of the mass operator $m_M$ in \cref{eq:dispersion} are obtained by projecting \cref{eq:massop} onto $\bu^M$ and inserting the decomposition, \cref{eq:G0}, of the bare Green's function. It is then a sum over the three waves $N$ into which the incident wave $M$ can be scattered,
\begin{equation}\label{eq:mM}
 m_M(k,\omega)=\sum_{N}m_{M\to N}(k,\omega),
\end{equation}
with
\begin{equation}\label{eq:channels}
\begin{split}
 m_{M\to N}(k,\omega)={}&\frac{k^2}{\rho^2}\int g^0_N(k^S)\,(k^S)^2\\
 &\times\mathrm{IP}_{M\to N}(\bp,\bs)\,W(\bk-\bk^S)\,\dd^3k^S,
\end{split}
\end{equation}
where the polarization of the scattered wave $N$ is $\bv^N$ and the inner product
\begin{equation}\label{eq:IP}
 \mathrm{IP}_{M\to N}(\bp,\bs)=\langle\delta c_{ijkl}\,\delta c_{\alpha\beta\gamma\delta}\rangle\,
 u^M_\beta u^M_k\,p_\alpha p_l\,v^N_\gamma v^N_j\,s_\delta s_i
\end{equation}
contracts the covariance with the polarizations and directions of the incident and scattered waves. For a longitudinal incident wave, the inner product depends only on the angle $\theta$ between $\bp$ and $\bs$. For a transverse incident wave, it also depends on the azimuth $\varphi$ of $\bs$ about $\bp$, measured from the polarization $\bu^M$. Because $W(\bk-\bk^S)$ depends on the direction of $\bs$ only through $\theta$, the azimuthal integration in \cref{eq:channels} acts on the inner product alone, and we write $\mathrm{IP}_{M\to N}(\theta)$ for its azimuthal average, which is the same for both transverse polarizations. For macroscopically isotropic media, it has the form \citep{Rokhlin2015,Sha2020}
\begin{equation}\label{eq:IPangle}
 \mathrm{IP}_{M\to N}(\theta)=A_{MN}+B_{MN}\cos^2\theta+C_{MN}\cos^4\theta,
\end{equation}
whose coefficients are quadratic invariants of the single-crystal elastic constants and are available in closed form for any crystal symmetry \citep{Sha2020}. Each term $m_{M\to N}$ in \cref{eq:channels} is a scattering channel. A longitudinal wave has the channels $\mathrm{L}\to\mathrm{L}$ and $\mathrm{L}\to\mathrm{T}$ (the latter combines the two transverse polarizations), and a transverse wave has the channels $\mathrm{T}\to\mathrm{T}$ and $\mathrm{T}\to\mathrm{L}$. The channel structure is the origin of the multiplicity examined in this paper.

The remaining integral in \cref{eq:channels} is evaluated by writing the bare Green's function in the Sokhotski--Plemelj form, $g^0_N=\PV[\omega^2-(k^S)^2V_{0N}^2]^{-1}-\jj\pi\delta(\omega^2-(k^S)^2V_{0N}^2)$, which separates the mass operator into a principal-value part and a residue part. Let $k_{0N}=\omega/V_{0N}$ be the wavenumber of the scattered wave in the reference medium, and let $\xi=k^S/k_{0N}$. With the azimuthal average taken in \cref{eq:IPangle}, each channel takes the form \citep{Sha2020}
\begin{equation}\label{eq:massop_final}
\begin{split}
 m_{M\to N}(k,\omega)={}&\frac{k^2k_{0N}^3}{\rho^2V_{0N}^2}\biggl[\PV\!\int_0^\infty\frac{\xi^4\,h_{M\to N}(k,\xi)}{1-\xi^2}\,\dd\xi\\
 &-\jj\frac{\pi}{2}\,h_{M\to N}(k,1)\biggr],
\end{split}
\end{equation}
where $h_{M\to N}(k,\xi)=\int_0^\pi\mathrm{IP}_{M\to N}(\theta)\,W_N(k,\xi,\theta)\sin\theta\,\dd\theta$ and
\begin{equation}\label{eq:WN}
\begin{split}
 &W_{N}(k,\xi,\theta)=2\pi\,W(k\bp-\xi k_{0N}\bs)\\
 &\quad=\frac{2a^3}{\pi\left[a^2\left(k^2-2k\xi k_{0N}\cos\theta+\xi^2k_{0N}^2\right)+1\right]^2}.
\end{split}
\end{equation}
The $\theta$ integrals are elementary for the angular form, \cref{eq:IPangle}, so that each channel reduces to a single principal-value integral over $\xi$, which we evaluate by Gauss--Legendre quadrature with the singular window at $\xi=1$ treated analytically. The dispersion function $D_M(k,\omega)$ can thus be evaluated for complex $k$, and its roots are found by a complex Newton iteration seeded from a scan of the $(f,k)$ plane and tracked continuously in frequency. This full evaluation of the FOSA is the theory against which the simulations are compared in Section~\ref{sec:results}. Its quasi-static velocity limit is the second-order static limit of \citet{Stanke1984}, which lies below the Voigt velocity by an amount that grows with the anisotropy. This limit is within 1\% of the self-consistent static estimate \citep{KubeDeJong2016} for transverse waves in iron, whereas it is 10\% above that estimate for longitudinal waves in the strongly anisotropic SrMgSn also studied here.

Finally, the two-point correlation function of the Voronoi microstructures simulated in this paper is measured numerically, and it is better represented by a sum of exponentials fitted to the measurement \citep{Sha2020},
\begin{equation}\label{eq:tpc}
\begin{gathered}
 w(r)=\sum_n w_n\exp(-r/a_n),\qquad \sum_n w_n=1,\\
 \aC=\int_0^\infty w(r)\,\dd r=\sum_n w_n a_n,
\end{gathered}
\end{equation}
which defines the correlation length $\aC$ and reduces to the single exponential for one term with $\aC=a$. Since the spectral density of \cref{eq:W} and thus the mass operator are linear in $w$, every expression above holds for the fitted function as the $w_n$-weighted sum over its terms with their own radii $a_n$. All theoretical results shown in this paper use this fit, while the single-exponential form is kept in Section~\ref{sec:ffa} to make the structure of the dispersion equation explicit.

\subsection{Far-field approximation and the three roots}\label{sec:ffa}

The dispersion equation, \cref{eq:dispersion}, is transcendental in $k$, and thus the number of its roots is not evident. The far-field approximation (FFA) of \citet{Rokhlin2015} makes the structure explicit. It rests on two simplifications. First, it takes the inner products out of the integrals in \cref{eq:channels} by the mean-value theorem, evaluating them in the forward direction for $N=M$ and as their angular average for $N\neq M$. Second, it replaces the bare Green's function by its far-field form, which keeps only the propagating $\exp(\jj k_{0N}r)/r$ term and discards the near-field $1/r^2$ and $1/r^3$ terms. The convolution in \cref{eq:channels} then reduces to a Fourier transform of the correlation function, and for a single exponential $w(r)=\exp(-r/a)$, each channel has the closed form
\begin{equation}\label{eq:ffa}
 m^\mathrm{FFA}_{M\to N}(k,\omega)=-\frac{4V_{0M}^2\,k^2\,k_{0N}^2\,Q_{M\to N}}{k^2-\left(k_{0N}+\jj/a\right)^2},
\end{equation}
where $Q_{M\to N}$ is a dimensionless scattering factor that measures the strength of the channel \citep{Rokhlin2015},
\begin{equation}\label{eq:QMN}
 Q_{M\to N}=\frac{\mathrm{Ip}_{M\to N}}{4\rho^2V_{0M}^2V_{0N}^2},
\end{equation}
with $\mathrm{Ip}_{M\to N}$ the inner product taken out of the integral by the first simplification. Substituting \cref{eq:ffa} into the dispersion equation, \cref{eq:dispersion}, results in the FFA model. The FFA loses the near-field terms that produce the static velocity shift, so its quasi-static velocity is the Voigt velocity. A constant shift equal to the second-order static correction restores the quasi-static limit \citep{Rokhlin2015}, which makes the FFA and FOSA roots agree closely at all frequencies \citep{Huang2021}.

Each channel in \cref{eq:ffa} is a rational function of $k^2$ with a simple pole at $k=\kappa_N$, where $\kappa_N=k_{0N}+\jj/a$. This pole has a clear physical meaning as a wave that travels at the velocity $V_{0N}$ of the scattered wave and decays over one correlation length. It describes the field scattered once into channel $N$, and its decay length reappears in Section~\ref{sec:wavefield} as the distance over which the incident plane wave is scattered out of the coherent field in the geometric regime. Because the two transverse polarizations form a single channel (Section~\ref{sec:fosa}), the mass operator of either polarization has exactly two distinct poles (one longitudinal and one transverse). Writing $\varepsilon_N=4V_{0M}^2k_{0N}^2Q_{M\to N}$ for the strength of channel $N$, the dispersion function becomes
\begin{equation}\label{eq:rational}
\begin{split}
 D_M(k,\omega)={}&\omega^2-V_{0M}^2k^2+\sum_{N=\mathrm{L},\mathrm{T}}\frac{\varepsilon_N\,k^2}{k^2-\kappa_N^2}\\
 ={}&\frac{P(k^2)}{\left(k^2-\kappa_\mathrm{L}^2\right)\left(k^2-\kappa_\mathrm{T}^2\right)},
\end{split}
\end{equation}
with the numerator the cubic polynomial in $k^2$
\begin{equation}\label{eq:cubic}
 P(k^2)=P_6k^6+P_4k^4+P_2k^2+P_0,
\end{equation}
where the coefficients are
\begin{equation}\label{eq:coeffs}
\begin{aligned}
 P_6&=-V_{0M}^2,\\
 P_4&=\omega^2+V_{0M}^2\left(\kappa_\mathrm{L}^2+\kappa_\mathrm{T}^2\right)+\varepsilon_\mathrm{L}+\varepsilon_\mathrm{T},\\
 P_2&=-\omega^2\left(\kappa_\mathrm{L}^2+\kappa_\mathrm{T}^2\right)-V_{0M}^2\kappa_\mathrm{L}^2\kappa_\mathrm{T}^2-\varepsilon_\mathrm{L}\kappa_\mathrm{T}^2-\varepsilon_\mathrm{T}\kappa_\mathrm{L}^2,\\
 P_0&=\omega^2\kappa_\mathrm{L}^2\kappa_\mathrm{T}^2.
\end{aligned}
\end{equation}
The dispersion equation $D_M=0$ is therefore equivalent to $P(k^2)=0$, a cubic in $k^2$ with three roots $K_i^2$, $i=1,2,3$. Each root gives a pair $\pm K_i$ of wavenumbers, and the one with positive real and imaginary parts ($K_i=k_i+\jj\alpha_i$) is a forward-propagating wave. The channel structure thus gives the dispersion equation three roots, with one root contributed by the bare pole of $g^0_M$ and one more by each distinct channel velocity. This count explains the earlier results. A scalar medium has a single channel and therefore two roots. They are the two terms given by \citet{Frisch1968} and the two coherent waves that \citet{Calvet2012} found for scalar waves. An isotropic elastic medium has two channels (longitudinal and transverse) and therefore three roots for each polarization. Two of them were found earlier as two peaks of the spectral function \citep{Calvet2012} and as two roots of the dispersion equation \citep{Rokhlin2015}, and the third was found later \citep{Huang2021}.

The physical origin of each root is clearest in the limit of weak scattering, where each root reduces to a known quantity of the unperturbed problem. As $Q_{M\to N}\to0$ and thus $\varepsilon_N\to0$ at fixed frequency, the cubic $P(k^2)$ of \cref{eq:cubic} factorizes into $(\omega^2-V_{0M}^2k^2)(k^2-\kappa_\mathrm{L}^2)(k^2-\kappa_\mathrm{T}^2)$, so one root tends to the bare wavenumber $k_{0M}$ of the reference medium and the other two tend to the channel poles $\kappa_\mathrm{L}$ and $\kappa_\mathrm{T}$. To first order in the scattering, the first root reproduces the Born attenuation of \citet[Eq.~27]{Rokhlin2015}. It is the effective wave of the classical description, whose attenuation rises as $\omega^4$ in the Rayleigh regime ($k_{0M}a\ll1$). The other two are heavily damped disturbances that travel near the longitudinal and transverse velocities and decay over about one correlation length. How close the actual roots lie to these limits depends on frequency. The coupling between the roots is measured by the channel terms of \cref{eq:rational} relative to the bare term $V_{0M}^2k^2$. It is of order $Q_{M\to N}(k_{0M}a)^2$ at low frequency and saturates at $Q_{M\to N}$ above $k_{0M}a\sim1$, so the roots stay close to their weak-scattering limits for $k_{0M}a\ll1$ and move away from them as frequency increases.

The opposite limit of $k_{0M}a\gg1$ therefore shows what the roots become once the coupling is fully developed. In this geometric regime, where the wavelength is shorter than the grains, the roots settle to limits that no longer depend on frequency. With the wavenumbers normalized by that of the incident wave as $\bar k=k/k_{0M}$ and $\bar k_N=k_{0N}/k_{0M}$, each channel term of \cref{eq:rational}, divided by $V_{0M}^2k_{0M}^2$, tends to $4Q_{M\to N}\bar k_N^2\bar k^2/(\bar k^2-\bar k_N^2)$, and the dispersion equation becomes the real cubic in $\bar k^2$
\begin{equation}\label{eq:geomcubic}
 1-\bar k^2+\sum_{N=\mathrm{L},\mathrm{T}}\frac{4Q_{M\to N}\,\bar k_N^2\bar k^2}{\bar k^2-\bar k_N^2}=0,
\end{equation}
whose roots give the phase velocities $V_{0M}/\bar k$ approached at high frequency. Two cases arise. In the first case, the coupling to the other channel is weak, and the bare root forms a pair with the root of its own channel ($N=M$). Keeping only that channel gives
\begin{equation}\label{eq:geompair}
 \bar k=\sqrt{1+Q_{M\to M}}\pm\sqrt{Q_{M\to M}},\quad 2\alpha a=\frac{1}{\sqrt{1+Q_{M\to M}}},
\end{equation}
where the two signs give the two roots of the pair. Their velocities $V_{0M}/\bar k$ follow from the real cubic of \cref{eq:geomcubic}, and their attenuation follows from its leading correction at large but finite $k_{0M}a$. The two roots of the pair are therefore positioned on either side of $V_{0M}$ at a relative distance of about $\sqrt{Q_{M\to M}}$ and have the same attenuation. A wave shorter than the grains samples the local wave speed of each grain, and $\sqrt{Q_{M\to M}}$ is the r.m.s.\ relative fluctuation of that speed \citep{Calvet2012}. The pair thus represents the slow and the fast parts of the distribution of local wave speeds, which Section~\ref{sec:ray} relates to the plane waves of the individual grains. In the second case, the coupling to the other channel is strong, which requires a large scattering factor $Q_{M\to N}$ and channel velocities that are close. The bare root is then joined by the root of that other channel instead of its own channel, and the strong coupling between the two pushes them far apart in velocity. In either case, the remaining root stays near its pole. Its attenuation is of order $1/a$ and is thus set by the correlation length instead of the frequency. Sections~\ref{sec:results_T} and \ref{sec:results_L} show that transverse waves in iron follow the first case and longitudinal waves in SrMgSn the second.

Throughout this paper, the roots are numbered so that root~1 is the root observed as the single mode at low frequency, root~2 is the root that joins it in the geometric regime and becomes observable, and root~3 is the remaining one, which is never observed. For both cases studied in Section~\ref{sec:results}, root~2 starts at the transverse channel pole and root~3 at the longitudinal one. Root~1 is the slower of the two observable transverse roots in iron and the faster of the two observable longitudinal roots in SrMgSn. The numbers on the figures follow this convention throughout.

The multi-term correlation function of \cref{eq:tpc} adds one pole per term to each channel and thus additional roots. These spurious roots are a consequence of representing $w(r)$ by a sum of exponentials. They stay close to their poles, are heavily damped and carry a negligible part of the spectral function, so they are not considered further.

\subsection{Spectral function and amplitude partition}\label{sec:spectral}

The roots of \cref{eq:dispersion} give velocities and attenuations but not amplitudes. The amplitude of each root is important because a root can be a mathematical solution without carrying a significant part of the coherent field. The amplitude is contained in the mean Green's function itself, and the natural quantity to examine is the spectral function \citep{Jing1992,Calvet2012},
\begin{equation}\label{eq:spectral}
 S_{\!M}(k,\omega)=-\Imag\,g_M(k,\omega)=-\Imag\frac{1}{D_M(k,\omega)},\qquad k\in\mathbb{R},
\end{equation}
which measures how strongly the mean field responds at each real wavenumber and frequency. A root appears in the $(f,k)$ plane as a ridge of $S_{\!M}$ centered on its wavenumber $k_i$ with a width set by its attenuation $\alpha_i$. \citet{Calvet2012} obtained the velocity and attenuation of coherent waves in polycrystals by fitting such ridges with Lorentzian peaks. However, a root too damped to form a ridge of its own cannot be fitted, and a fit does not reveal how the weight of the spectral function is divided between overlapping roots.

To address both limitations, we propose a method to calculate the amplitude of each root directly from the residue of the mean Green's function at that root. The mean Green's function is decomposed into a sum of terms, with one term for each root weighted by its residue. In the far-field approximation, this decomposition is exact without remainder because the mean Green's function is the rational function $1/D_M=(k^2-\kappa_\mathrm{L}^2)(k^2-\kappa_\mathrm{T}^2)/P(k^2)$ of $k^2$. Its numerator is of lower degree than its denominator, so it has the partial-fraction expansion
\begin{equation}\label{eq:partial}
\begin{gathered}
 \frac{1}{D_M(k,\omega)}=\sum_{i=1}^3\frac{R_i}{k^2-K_i^2},\\
 R_i=\left.\frac{1}{\partial D_M/\partial(k^2)}\right|_{K_i}
 =\left.\frac{2K_i}{\partial D_M/\partial k}\right|_{K_i}\!\!,
\end{gathered}
\end{equation}
in which $R_i$ is the residue at the pole $K_i^2$ of root $i$. Because the numerator has degree two and the denominator degree three, the residues satisfy a sum rule in which their sum depends only on the reference medium:
\begin{equation}\label{eq:sumrule}
 \sum_{i=1}^3R_i=\frac{1}{P_6}=-\frac{1}{V_{0M}^2}.
\end{equation}
Taking the imaginary part of \cref{eq:partial} for real $k$ and writing $K_i^2=k_i^2-\alpha_i^2+\jj\Gamma_i$ with $\Gamma_i=2k_i\alpha_i$, the spectral function splits into one term per root,
\begin{equation}
 S_{\!M}(k,\omega)=\sum_{i=1}^3\sigma_i(k),
\end{equation}
with
\begin{equation}\label{eq:modeterm}
\begin{gathered}
 \sigma_i(k)=\frac{A_i\Gamma_i-B_i\,\Delta_i}{\Delta_i^2+\Gamma_i^2},\qquad
 \Delta_i=k^2-k_i^2+\alpha_i^2,\\
 A_i=-\Real R_i,\qquad B_i=\Imag R_i.
\end{gathered}
\end{equation}
The first part of $\sigma_i$ is a Lorentzian in $k^2$ that is centered on $\Delta_i=0$ (i.e.\ $k^2=k_i^2-\alpha_i^2$) and has a half-width $\Gamma_i$. It is precisely the peak fitted by \citet[Eq.~(36)]{Calvet2012}, and their fitted amplitude equals $A_i\Gamma_i$ here. The second part is antisymmetric about the same center and is present whenever the residue is complex, which is generally the case for a damped root. The antisymmetric part integrates to zero in the principal-value sense, so the weight of root $i$ in the spectral function is the integral of the symmetric part, which is approximately $\int\sigma_i\,\dd(k^2)=\pi A_i$ for a narrow peak whose half-width $\Gamma_i$ is small compared with $k_i^2$ (i.e.\ $2\alpha_i\ll k_i$). Taking the real and imaginary parts of the sum rule, \cref{eq:sumrule}, gives
\begin{equation}\label{eq:sumruleAB}
 \sum_{i=1}^3A_i=\frac{1}{V_{0M}^2},\qquad \sum_{i=1}^3B_i=0.
\end{equation}
The total spectral weight at any frequency is thus fixed for the given reference medium, which means that scattering neither creates nor destroys it but only redistributes it among the roots. We therefore call the fraction
\begin{equation}\label{eq:partition}
 \eta_i=A_iV_{0M}^2
\end{equation}
the amplitude partition of root $i$. It is the fraction of the spectral weight that the root carries at that frequency, and the partitions of the three roots add to unity.

The residue method has two advantages. First, since the roots are known, the residues are obtained from the derivative of the dispersion function at the pole without any fitting. \Cref{eq:partial} thus applies to a root of any attenuation, including one so damped that it never forms a peak. Second, it separates the two factors that decide whether a root is observed, which are its weight and its attenuation. The height of its ridge in the $(f,k)$ plane follows from \cref{eq:modeterm} as
\begin{equation}\label{eq:height}
 H_i=\sigma_i\big|_{\Delta_i=0}=\frac{A_i}{\Gamma_i}=\frac{A_i}{2k_i\alpha_i}.
\end{equation}
A small partition and a large attenuation both lower the ridge, and their effects multiply. \Cref{eq:height} thus quantifies both factors, whereas our previous work considered the attenuation alone \citep{Huang2026prl}.

\begin{figure*}[!t]
 \centering
 \includegraphics[width=0.8\textwidth]{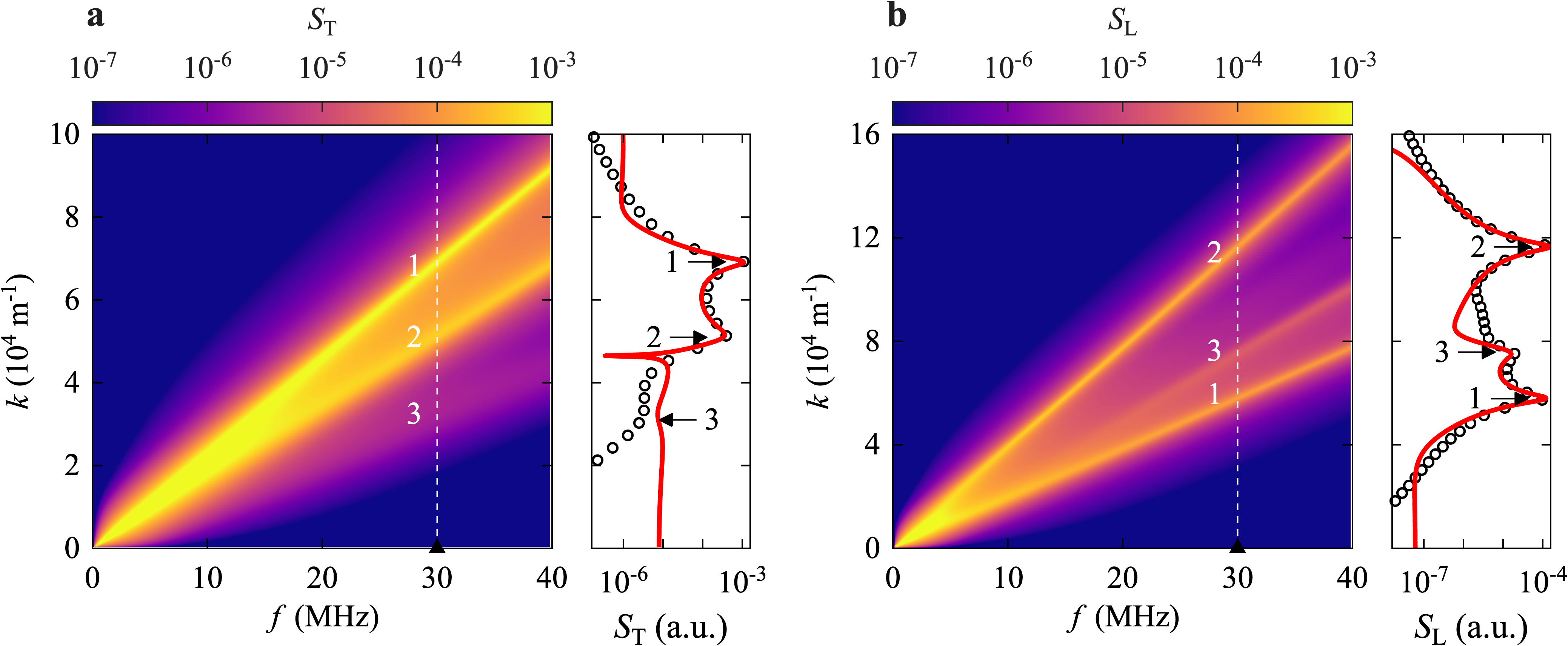}
 \caption{Spectral function of the mean Green's function and its decomposition into the three roots. The maps show $S_{\!M}(k,\omega)=-\Imag[1/D_M]$ from the first-order smoothing approximation for (a) transverse waves in polycrystalline iron and (b) longitudinal waves in polycrystalline SrMgSn, on a logarithmic color scale spanning four decades. The numbers label the three roots of the dispersion equation, \cref{eq:dispersion}. The dashed line marks $f=30$~MHz, and the panel to the right of each map shows the section of the spectral function at that frequency: points are the exact spectral function, \cref{eq:spectral}, and the red line is the sum, \cref{eq:modeterm}, of the three root terms built from the residues at the roots, with the arrows pointing at the peaks of the three roots.}
 \label{fig:spectral}
\end{figure*}

In the full FOSA theory, the dispersion function is not rational, and the pole expansion of \cref{eq:partial} is an approximation whose accuracy has to be checked. We evaluate the residues numerically from \cref{eq:partial}, with $\partial D_M/\partial k$ obtained by a central difference of the dispersion function at each root, and compare the sum of the three root terms with the spectral function computed directly from \cref{eq:spectral}. Figure~\ref{fig:spectral} shows the result for the two cases simulated in this paper, which are transverse waves in iron and longitudinal waves in SrMgSn (the materials are specified in Section~\ref{sec:model}). In both materials, a single bright ridge at low frequency divides into two in the geometric regime, and a third, much fainter ridge runs between or beside them at all frequencies. The profiles at 30~MHz show the exact spectral function as points and the sum of the three root terms as the red line. The sum reproduces the position, height and width of every peak, including that of the faint root~3 in SrMgSn, which confirms that the three residues capture every root of the mean Green's function. The sum differs from the exact function only between the peaks, where no root dominates and the antisymmetric parts of neighboring roots partly cancel. Integrated over wavenumber, the three residues account for the exact spectral weight to within about 15\% at every frequency examined in both materials. The remainder lies outside the three pole terms and belongs to the heavily damped additional roots of the multi-term correlation function (Section~\ref{sec:ffa}) and to the part of the full dispersion function that is not rational. The sum rule of \cref{eq:sumruleAB} holds approximately with $V_{0M}$ replaced by the quasi-static velocity of the FOSA, which is the velocity to which root~1 tends as $\omega\to0$.

The residue method therefore carries over to the FOSA. We use it to obtain the amplitude partitions and ridge heights of the three roots in Section~\ref{sec:partition}, with the partitions normalized so that they add to unity.

\section{Simulation}\label{sec:simulation}

This section describes the finite-element models (Section~\ref{sec:model}), the wavefields they produce and the mechanism of multiplicity they reveal (Section~\ref{sec:wavefield}), and the extraction of the coherent modes from them (Section~\ref{sec:processing}).

\subsection{Finite-element models}\label{sec:model}

\begin{figure*}[!t]
 \centering
 \includegraphics[width=0.8\textwidth]{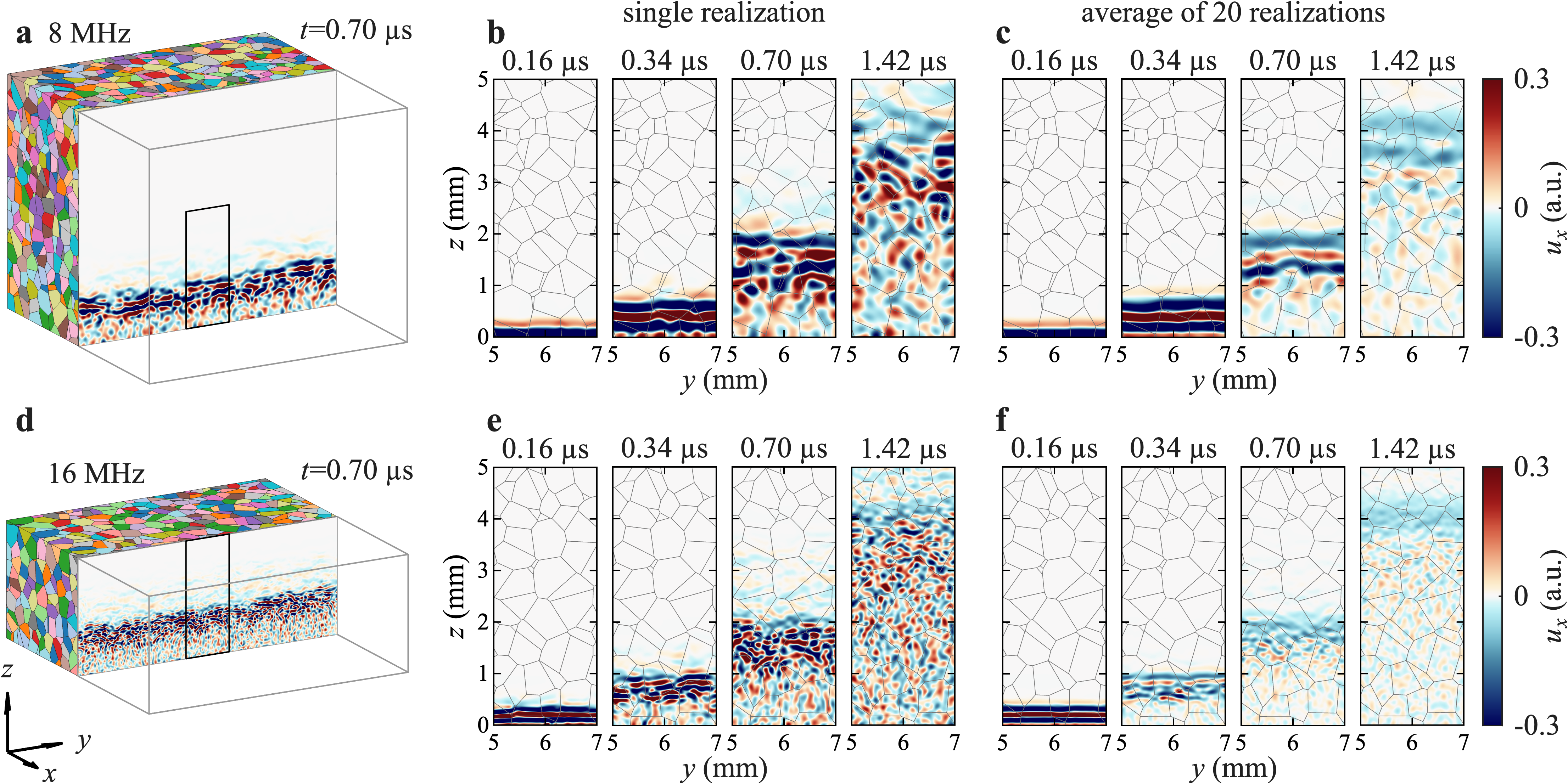}
 \caption{Wavefield of a plane transverse pulse in polycrystalline iron. The pulse is excited on the plane $z=0$ of a statistically isotropic aggregate, with polarization along $x$, at 8~MHz (top row) and 16~MHz (bottom row) center frequency. (a,~d)~Computational domains cut open at $x=6$~mm, with the grains colored arbitrarily and the removed half outlined in gray; color on the cut face is the displacement $u_x$ at $t=0.70~\us$ in a single realization, and the black rectangle marks the window of the snapshots. The 8~MHz domain is $12\times12\times10$~mm$^3$ and the 16~MHz domain $12\times12\times5$~mm$^3$. (b,~e)~$u_x$ on the cut plane in a single realization at the four times shown, with the grain boundaries on that plane overlaid. (c,~f)~The same for the average over 20 realizations with re-randomized grain orientations. In each row, the color scale saturates at $\pm0.3$ of the peak amplitude of the averaged coherent field of that row, in arbitrary units.}
 \label{fig:wavefield}
\end{figure*}

The polycrystals are modeled explicitly at the grain scale, and the elastic wave equation is solved in three dimensions with the GPU-accelerated time-domain finite-element solver Pogo \citep{Huthwaite2014}, following the methodology that we established and validated in earlier work \citep{VanPamel2017,Huang2020a,Huang2022a,Huang2022b}. The simulations thus invoke no scattering theory, and they test the theory of Section~\ref{sec:theory} independently. Two cases are studied, which are transverse waves in iron (a cubic material of common anisotropy) and longitudinal waves in SrMgSn (a cubic material of extreme anisotropy). The two cases share the same simulated grain structures. Two cuboid computational domains of $12\times12$~mm$^2$ cross section are used. One is 5~mm long and contains 5760 grains, and the other is 10~mm long and contains 11520 grains. The grains are generated by Voronoi tessellation with Neper \citep{Quey2011}, are equiaxed with mean size $d=0.5$~mm, and have the two-point correlation function of \cref{eq:tpc} with correlation length $\aC=0.23$~mm \citep{Sha2020}. Each grain is assigned a crystallographic orientation drawn uniformly at random, which makes the aggregate statistically isotropic. The grains are lossless (no intrinsic dissipation), so all attenuation of the coherent wave arises from scattering. The domains are discretized into uniform structured meshes of eight-node hexahedral elements. The element size is 12.5~$\upmu$m for the 5~mm domain (369 million elements) and 25~$\upmu$m for the 10~mm domain (92 million elements). Figure~\ref{fig:wavefield}(a) and (d) show the two domains cut open at $x=6$~mm. In every simulation, a three-cycle toneburst force (a short sinusoidal pulse of three cycles) is applied uniformly over the face $z=0$. Twenty realizations of each simulation are computed with the same tessellation but re-randomized grain orientations, which we found sufficient for the coherent field to converge. The displacement components orthogonal to the excited polarization are also recorded and are examined in Appendix~\ref{sec:offpol}.

For the transverse case, the grains are of iron with elastic constants $c_{11}=219.2$, $c_{12}=136.8$ and $c_{44}=109.2$~GPa, density $\rho=7860$~kg/m$^3$, Zener anisotropy index $A=2.65$, universal anisotropy index \citep{Ranganathan2008} $A^U=1.23$ and log-Euclidean anisotropy index \citep{Kube2016} $A^L=0.49$. Its Voigt velocities are $\vL=5900$~m/s and $\vT=3230$~m/s. The excitation force is directed along $x$, and the four lateral surfaces have their $y$ and $z$ motions restrained, which enforces a plane transverse wave propagating along $z$. The simulations are run at center frequencies of 8~MHz on the 10~mm domain and 16~MHz on the 5~mm domain, corresponding to Voigt wavelengths $\lambda_\mathrm{0T}=0.40$ and 0.20~mm and normalized frequencies $2\kT\aC=7.2$ and 14.4, with $\kT=2\pi f/\vT$. Both thus have 16 elements per Voigt wavelength at the center frequency, which gives a numerical error of about 0.1\% in the phase velocity \citep{Huang2020a}.

The longitudinal case differs from the transverse one in three respects. First, the grains are of SrMgSn, with $c_{11}=7$, $c_{12}=4$ and $c_{44}=25$~GPa, $\rho=3472.8$~kg/m$^3$, $A=16.67$, $A^U=17.67$ and $A^L=3.38$, and Voigt velocities $\vL=2726$~m/s and $\vT=2119$~m/s. At a given frequency, the longitudinal wavelength is longer than the transverse one, and the strong anisotropy of SrMgSn brings the geometric regime of the longitudinal wave within computationally accessible frequencies \citep{Huang2022b}. Second, the excitation force is directed along $z$, and symmetry boundary conditions are applied on the lateral surfaces to enforce a plane longitudinal wave propagating along $z$. Third, the simulations are run at 2 and 4~MHz on the 10~mm domain and at 8~MHz on the 5~mm domain, which correspond to $\lambda_\mathrm{0L}=1.36$, 0.68 and 0.34~mm and $2\kL\aC=2.1$, 4.3 and 8.5. The longer wavelengths give at least 27 elements per Voigt wavelength. The single-crystal velocities of SrMgSn range from 1420 to 3322~m/s for quasi-longitudinal waves and from 657 to 2683~m/s for quasi-transverse waves, and this spread will matter in Section~\ref{sec:results_L}.

\subsection{Wavefield and the mechanism of multiplicity}\label{sec:wavefield}

Before reducing the wavefield to the coherent field, it is instructive to look at it directly because it shows the mechanism that produces multiple modes. Figure~\ref{fig:wavefield} shows the transverse displacement $u_x$ in iron at 8 and 16~MHz, in a single realization and after averaging over the twenty realizations. Only the transverse case is shown because the wavefields of the longitudinal case in SrMgSn show the same behavior.

We first consider the 8~MHz excitation with $2\kT\aC=7.2$, which is below the geometric regime. In the single realization of panel (b), the wavefront is smooth at $t=0.16~\us$, when it has traveled about 0.5~mm, and it is still intact at $0.34~\us$. At $0.70~\us$ (about 2~mm from the source), the wavefront remains recognizable, although the grains have distorted it and speckle (a random pattern of the grain size) has built up behind it. The speckle takes over only by $1.42~\us$. The average in panel (c) removes the speckle and leaves a single coherent wavefront, which weakens as it travels but is still visible at $1.42~\us$. At this frequency, the incident wave is thus converted gradually into the incoherent field, and the coherent field is a single wave, the effective wave of the classical description (root~1 of Section~\ref{sec:ffa}).

The picture changes at 16~MHz ($2\kT\aC=14.4$) in the geometric regime. In the single realization of panel (e), the wavefront is smooth at $0.16~\us$ but is already broken up on the scale of the grains by $0.34~\us$ (after about 1~mm), and the field from $0.70~\us$ onward is speckle with no visible trace of the incident plane wave. The snapshots cannot resolve the break-up more finely because the three-cycle pulse (about $0.19~\us$ long) is still being launched at earlier times. The theory places the extinction of the incident plane wave within about one correlation length, which is the decay length of the channel poles identified in Section~\ref{sec:ffa} ($\Imag\kappa_N=1/a$), and the snapshots are consistent with it. However, the average in panel (f) still shows a coherent wavefront at $0.70~\us$ (about 2~mm from the source) and a weak one at $1.42~\us$. This coherent wave at depth cannot be the incident wave because the latter has been extinguished. It is rebuilt from the scattered field by forward multiple scattering and inherits the properties of the channels through which the scattering proceeds. Since each scattering channel contributes a root besides that of the incident wave (Section~\ref{sec:ffa}), the coherent field rebuilt by scattering can carry more than one mode, and this is the physical content of the three roots. Section~\ref{sec:ray} uses a grain velocity estimate to show how the coherent field forms in one layer of grains from the plane waves of the individual crystals. The snapshots do not separate these modes, which are instead resolved in the coherent field and its spectrum in Section~\ref{sec:processing}.

\subsection{Coherent field and its spectral analysis}\label{sec:processing}

Now the simulated wavefields are reduced to measured modes in four steps that are the same for both transverse and longitudinal cases: the coherent field is extracted as a two-dimensional record, high-pass filtered, transformed into the frequency--wavenumber plane and fitted there with Lorentzian line shapes. We first describe the steps for the transverse case, and then the differences for the longitudinal case.

\begin{figure*}[!t]
 \centering
 \includegraphics[width=0.8\textwidth]{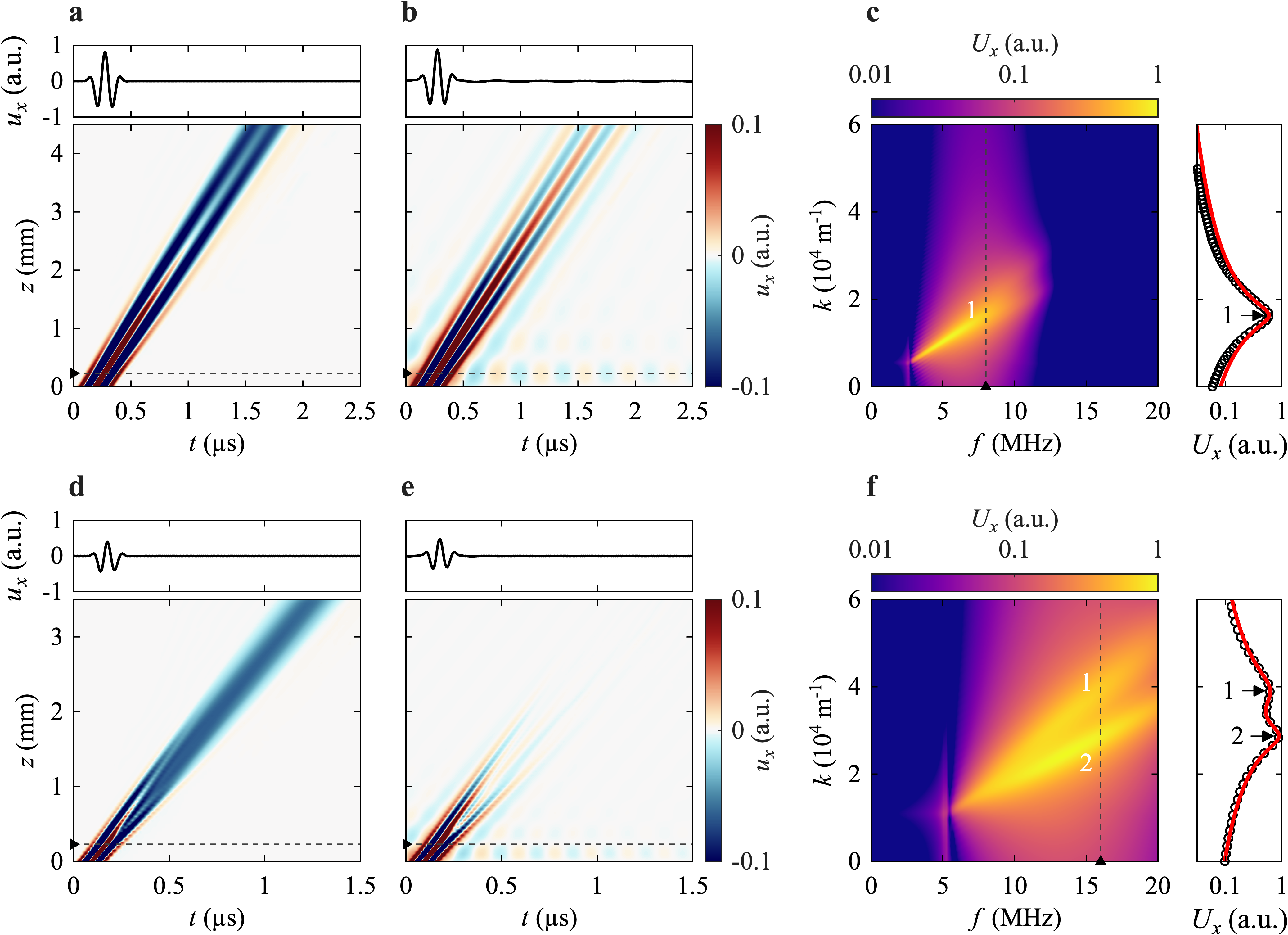}
 \caption{Coherent transverse field in iron and its spectral analysis. Results are shown for 8~MHz (top row) and 16~MHz (bottom row) center frequency. (a,~d)~Coherent field $u_x(t,z)$, averaged over the $x$--$y$ cross section at each depth $z$ and over 20 realizations, normalized to its peak; the strip above shows the waveform at $z=0.25~\mathrm{mm}\simeq\aC$, marked by the dashed line and arrow in the map. (b,~e)~The same after high-pass filtering above one third of the center frequency. (c,~f)~Magnitude of the two-dimensional Fourier transform of (b) and (e), normalized to its maximum and shown on a logarithmic color scale; points at right are the section at the center frequency, marked by the dashed line and arrow in the map, and the red line is the Lorentzian fit, \cref{eq:lorentz}, with the numbers labeling the modes according to the convention of Section~\ref{sec:ffa}.}
 \label{fig:processing_T}
\end{figure*}

For the transverse case, the displacement $u_x$ is averaged over the $x$--$y$ cross section at each depth $z$ and then over the twenty realizations, which turns the wavefield of Section~\ref{sec:wavefield} into its coherent part. The resulting trace at each depth is gated in time to the coherent pulse, which excludes late scattered arrivals. Along $z$, the traces are taken every 0.05~mm from the excited face to $z=4.5$~mm in the 8~MHz simulation and to 3.5~mm in the 16~MHz simulation, which keeps them clear of the reflection from the back face. Each simulation thus yields a two-dimensional record of the coherent field $u_x(t,z)$ as shown in Fig.~\ref{fig:processing_T}(a,d).

\begin{figure*}[!t]
 \centering
 \includegraphics[width=0.8\textwidth]{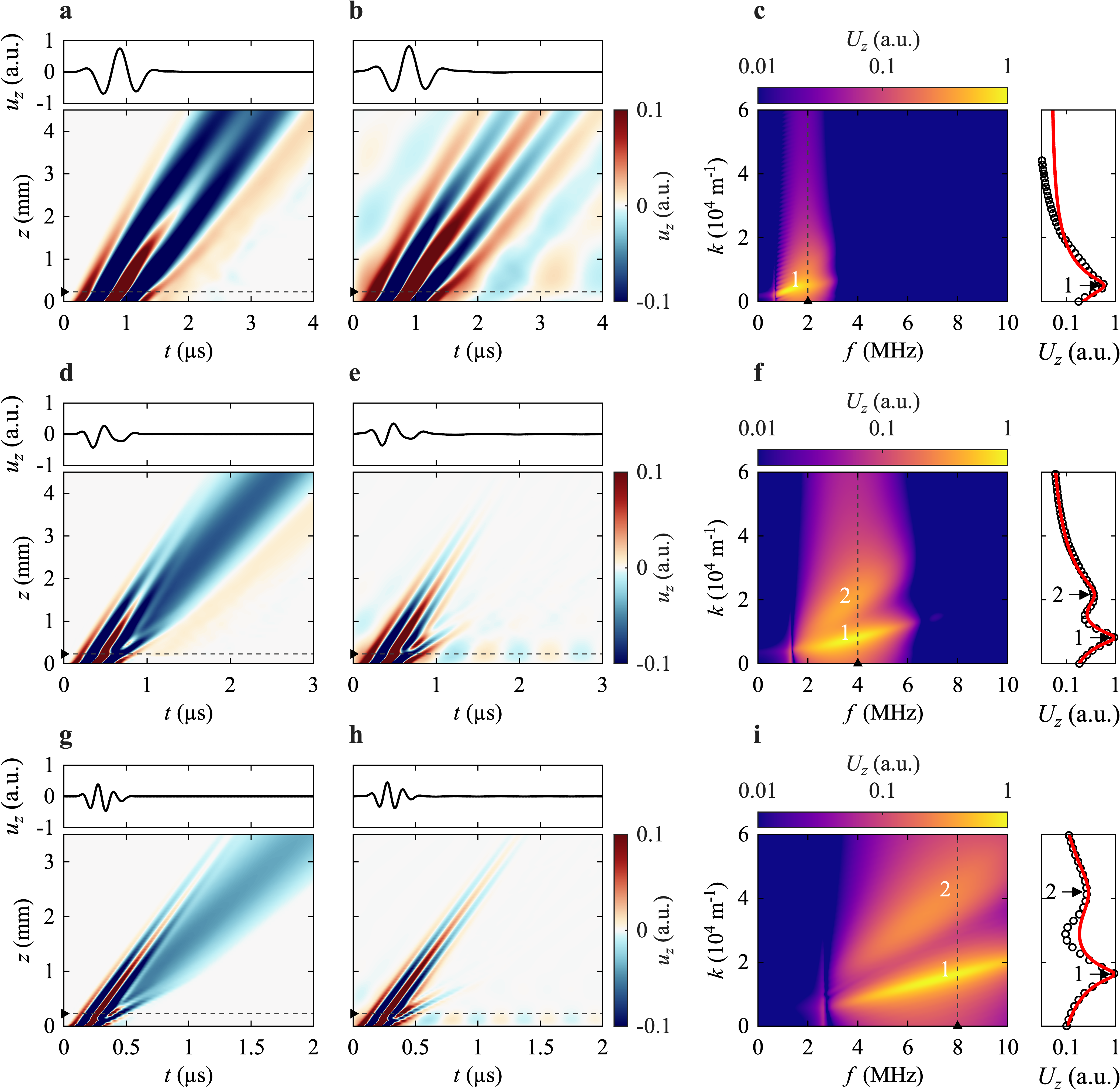}
 \caption{Coherent longitudinal field in SrMgSn and its spectral analysis. Same as Fig.~\ref{fig:processing_T}, for $u_z(t,z)$ at 2~MHz (top row), 4~MHz (middle row) and 8~MHz (bottom row) center frequency.}
 \label{fig:processing_L}
\end{figure*}

At 8~MHz ($2\kT\aC=7.2$), the record shows a single streak in Fig.~\ref{fig:processing_T}(a). It is then high-pass filtered to remove the low-frequency tail of the toneburst spectrum using a zero-phase, minimum-order infinite-impulse-response (IIR) high-pass filter of 60~dB stopband attenuation. Its passband edge is set at $f_\mathrm{c}(N-2)/N$, which is the lower bandwidth limit of an $N$-cycle toneburst of center frequency $f_\mathrm{c}$. The filtered record in Fig.~\ref{fig:processing_T}(b) still shows a single streak, so the coherent field carries one mode at this frequency. At 16~MHz ($2\kT\aC=14.4$), the streak near the source splits beyond $z\simeq\aC$ into two streaks of different slopes in Fig.~\ref{fig:processing_T}(d), which is consistent with the extinction of the incident wave within one correlation length and the rebuilding of the coherent field by scattering (Section~\ref{sec:wavefield}). Because attenuation grows with frequency, the split streaks are masked beyond $z\simeq5\aC$ by the surviving low-frequency components of the toneburst, but they emerge clearly after the same filtering in Fig.~\ref{fig:processing_T}(e). The spectra below are computed from the filtered records.

A two-dimensional Fourier transform \citep{Alleyne1991} then maps the filtered record into the frequency--wavenumber plane as $U_x(f,k)$. Its magnitude in Fig.~\ref{fig:processing_T}(c,f) has ridges of high amplitude that trace the dispersion relations of the coherent modes. A single ridge at 8~MHz becomes two ridges of distinctly different slopes and widths at 16~MHz, which are two coherent modes with their own phase velocities and attenuations. The two modes are unambiguously resolved, and this constitutes a direct observation of multiple coherent modes in a single-phase random medium.

To measure each mode, we fit the spectrum at each frequency. A mode with the complex wavenumber $k_i+\jj\alpha_i$ propagates as $\exp(\jj k_iz-\alpha_iz)$, and the magnitude squared of its one-sided spatial transform is a Lorentzian centered on $k_i$ with half-width $\alpha_i$. We therefore fit
\begin{equation}\label{eq:lorentz}
 F(k)=\sum_{i=1}^{2}\frac{b_i}{(k-k_i)^2+\alpha_i^2}+b_0,
\end{equation}
to the power spectrum at each frequency \citep{Huang2026jasa}. Two Lorentzians are used because the theory of Section~\ref{sec:spectral} predicts the ridge of the third root to lie about two orders of magnitude or more below the others. A fit with three Lorentzians returns the same two modes, as shown by \citet{Huang2026jasa} together with the fitting procedure. Two modes closer than about their half-widths cannot be resolved, and only one is then reported. The lowest frequency at which the second mode appears is thus set by this resolution limit instead of the physics. The fits are illustrated by the profiles beside the spectra and yield the measured phase velocities $V_i=\omega/k_i$ and attenuation coefficients $\alpha_i$ that are compared with theory in Section~\ref{sec:results}. The transverse modes are fitted from 2.5 to 12~MHz in the 8~MHz simulation and from 5 to 22~MHz in the 16~MHz simulation, and the figures show only the parts of these ranges that lie within the bandwidth of each toneburst.

The fitted amplitudes $b_i$ also give the partition $\eta^\mathrm{s}_i$ of each mode, which is the fraction of the spectral weight that it carries. Because the Lorentzian of each mode in \cref{eq:lorentz} is the squared magnitude of its spatial transform, its numerator $b_i$ is the squared displacement amplitude of the mode at the excited face. With the plane-force excitation and the boundary conditions of the models, the domain behaves as an unbounded polycrystal, in which a plane force excites each root of the mean Green's function, \cref{eq:partial}, with an amplitude proportional to $R_i/K_i$. The magnitude of the residue of mode~$i$ is therefore proportional to $k_i\sqrt{b_i}$ (since $\alpha_i\ll k_i$), which gives
\begin{equation}\label{eq:simpartition}
 \eta^\mathrm{s}_i=\frac{k_i\sqrt{b_i}}{\sum_{j=1}^{2}k_j\sqrt{b_j}}.
\end{equation}
This partition uses $|R_i|$ in place of $-\Real R_i$, so it equals the partition $\eta_i$ of \cref{eq:partition} where the residues are nearly real.

The accuracy of the fitted attenuations and amplitudes is limited by two effects, which are the finite length of the record along $z$ and the interference between modes neglected by \cref{eq:lorentz}. The first is negligible for strongly attenuated modes that decay well within the record, and the second is negligible for modes that are well separated in wavenumber. Both are analyzed in detail by \citet{Huang2026jasa}. Neither appreciably affects the centers of the peaks and hence the phase velocities. The modes do not depend on the average over realizations either. The cross-section average spans about 24 grains in each direction and already resolves both modes in every single realization, and the fitted velocities at 16~MHz vary between realizations by less than 1\% and the attenuation coefficients by about 6\%.

The longitudinal case is processed in the same way with three differences. First, the record is of the displacement along the excitation $u_z(t,z)$ as shown in Fig.~\ref{fig:processing_L}(a,d,g). Second, it extends to $z=4.5$~mm in the 2 and 4~MHz simulations and to 3.5~mm in the 8~MHz simulation. Third, the modes are fitted from 1 to 3, 1.5 to 7 and 3 to 13~MHz in the 2, 4 and 8~MHz simulations. The records show the same features as in the transverse case, with a single streak at 2~MHz ($2\kL\aC=2.1$) and a streak that splits beyond $z\simeq\aC$ at 4 and 8~MHz, as seen most clearly in the filtered records of Fig.~\ref{fig:processing_L}(e,h). Correspondingly, the single ridge of the 2~MHz spectrum becomes two ridges at 4 and 8~MHz in Fig.~\ref{fig:processing_L}(c,f,i).

\section{Grain velocity estimate}\label{sec:ray}

The theory of Section~\ref{sec:theory} deals with the polycrystal on the macroscopic scale, as a homogeneous reference medium perturbed by fluctuations. Here we develop a microscopic treatment on the scale of the grains in the geometric regime, where the wavelength is shorter than the grains and propagation can be pictured as rays that cross each grain as the plane waves of that crystal \citep{Rokhlin1992}. On this basis, we estimate the velocities and the partitions of the coherent modes using only the single-crystal elastic constants. The estimate follows the plane waves that the excitation launches in the individual grains and averages their slownesses over the grain orientations. We call the velocities obtained from these averages the grain velocities. The estimate provides an independent test of the theory and the simulations, and it shows how the coherent modes are formed from these waves.

\begin{figure*}[!t]
 \centering
 \includegraphics[width=0.8\textwidth]{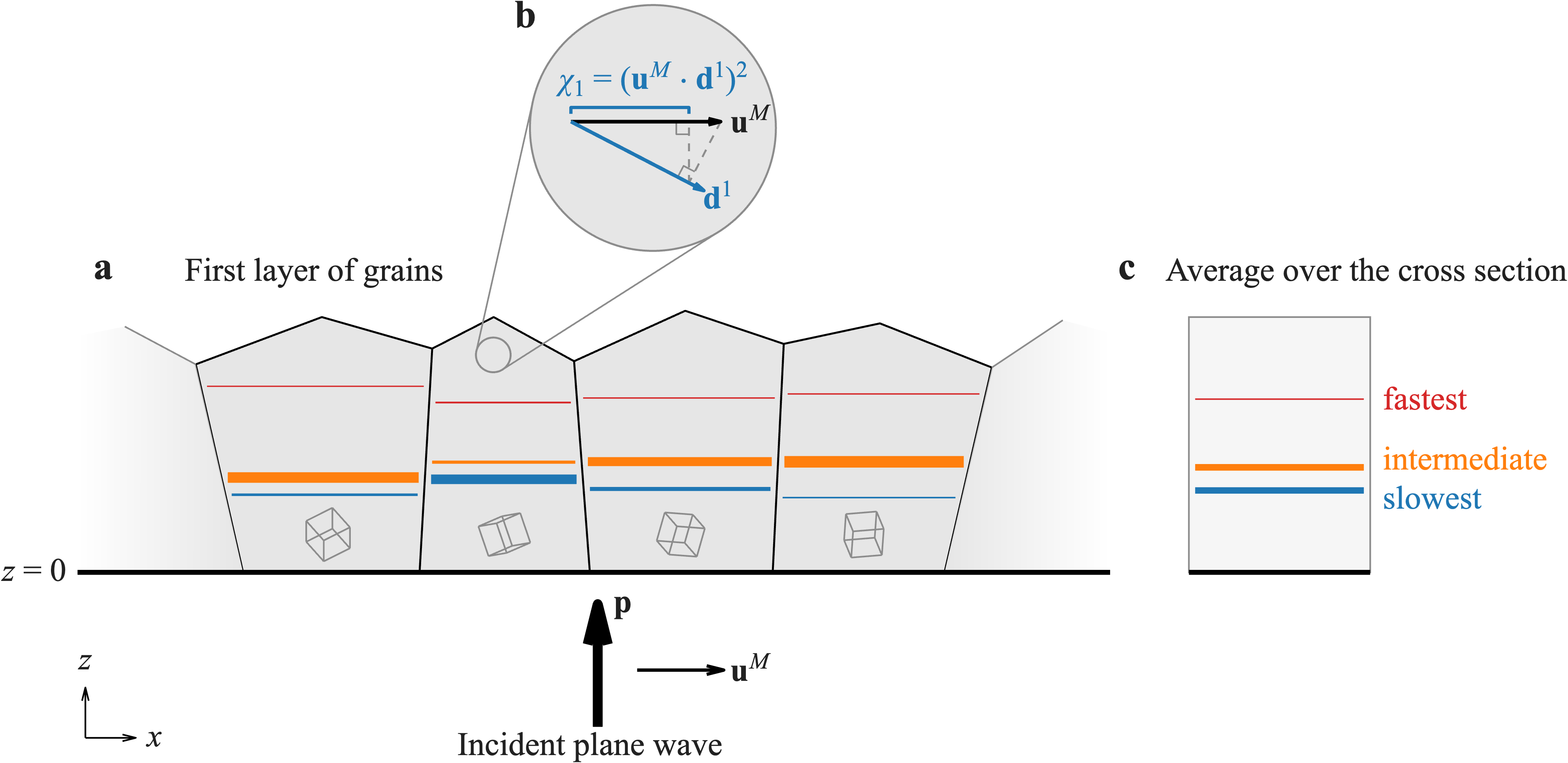}
 \caption{Schematic of the grain velocity estimate for a transverse excitation in iron. (a)~First layer of grains under a plane wave with wave normal $\bp$ and polarization $\bu^M$, incident on the face $z=0$. Each grain, whose crystal orientation is shown by the cube, launches its three crystal waves along $\bp$ at its own phase velocities, and the wavefronts are shown at one instant with line widths that grow with the weight $\chi_m$, \cref{eq:rayweight}. (b)~One grain seen along $\bp$, where $\bu^M$ is resolved onto the polarization $\mathbf{d}^1$ of the slowest wave and back onto $\bu^M$ to give $\chi_1$. (c)~Average over the cross section, with wavefronts that advance at the grain velocities $\bar V_m$, \cref{eq:rayvel}, and line widths that grow with the shares $\langle\chi_m\rangle$. The phase velocities and polarizations are computed from the single-crystal constants of iron for four random orientations.}
 \label{fig:schematic}
\end{figure*}

In this estimate, we consider only the first layer of grains behind the excited face, under a plane wave incident along the unit vector $\bp$ with unit polarization $\bu^M$, as shown in Fig.~\ref{fig:schematic}(a) for a transverse excitation. In a grain with elastic tensor $c_{ijkl}$, the incident plane wave excites three plane waves that share its wave normal $\bp$ and are given by the Christoffel equation
\begin{equation}\label{eq:christoffel}
 \left[c_{ijkl}\,p_jp_l-\rho\left(V^\mathrm{c}_m\right)^2\delta_{ik}\right]d^m_k=0,\qquad m=1,2,3,
\end{equation}
with phase velocities $V^\mathrm{c}_m$, numbered in order of increasing velocity, and orthonormal polarizations $\mathbf{d}^m$. We refer to these crystal waves as the slowest, the intermediate and the fastest. 

Because the phase velocities change from grain to grain, the wavefronts launched in neighboring grains are staggered (Fig.~\ref{fig:schematic}(a)). Each wavefront advances along $\bp$ at the phase velocity of its wave. In an anisotropic crystal, the energy of the wave travels at the group velocity $\mathbf{V}^\mathrm{g}_m$ in a direction that deviates from $\bp$, but it keeps pace with the wavefront along $\bp$ because the component of the group velocity along the wave normal is the phase velocity, i.e.\ $\mathbf{V}^\mathrm{g}_m\cdot\bp=V^\mathrm{c}_m$ \citep{Auld1973}. Only the progress along $\bp$ matters, because the coherent field is the average over the cross section and thus depends on the depth $z$ alone. Its phase is therefore governed by the slowness $1/V^\mathrm{c}_m$, the reciprocal of the phase velocity, and the estimate is built on these slownesses.

The phase velocities set the phases of the crystal waves, and their polarizations set the amplitudes. Within each grain, the incident displacement is decomposed into the three crystal waves, and its component along $\bu^M$ at a depth $z$ is
\begin{equation}\label{eq:raygrain}
 u(z)=\sum_{m=1}^3\chi_m\exp\!\left(\jj\omega z/V^\mathrm{c}_m\right),
\end{equation}
where
\begin{equation}\label{eq:rayweight}
 \chi_m=\left(\bu^M\cdot\mathbf{d}^m\right)^2,\qquad \sum_{m=1}^3\chi_m=1,
\end{equation}
is the weight of wave $m$. As Fig.~\ref{fig:schematic}(b) shows for the slowest wave, the weight is a square because the projection $\bu^M\cdot\mathbf{d}^m$ enters twice, first when the incident displacement along $\bu^M$ is resolved onto $\mathbf{d}^m$ to give the amplitude of wave $m$, and again when the displacement of that wave is resolved back onto $\bu^M$, which is the polarization of the coherent field.

Within the first layer of grains, the coherent field is the average of \cref{eq:raygrain} over the grain orientations,
\begin{equation}\label{eq:rayfield}
 \langle u\rangle(z)=\sum_{m=1}^3\big\langle\chi_m\exp\!\left(\jj\omega z/V^\mathrm{c}_m\right)\big\rangle,
\end{equation}
so each crystal wave contributes a band of slownesses $1/V^\mathrm{c}_m$ weighted by $\chi_m$. Each band is characterized by three numbers, and Fig.~\ref{fig:schematic}(c) illustrates the first two. The first is the velocity obtained from its mean slowness,
\begin{equation}\label{eq:rayvel}
 \bar V_m=\frac{\langle\chi_m\rangle}{\langle\chi_m/V^\mathrm{c}_m\rangle},
\end{equation}
which is the grain velocity and sets the phase of its contribution. The slowness is averaged instead of the velocity because the phase accumulated over a depth $z$, $\omega z/V^\mathrm{c}_m$, is linear in the slowness. The second number is its share $\langle\chi_m\rangle$ of the excitation. Because the share is a fraction of the incident displacement and the amplitude partition $\eta_i$ of Section~\ref{sec:spectral} is a fraction of the spectral weight, the share must be converted before the two are compared. The Green's function of the grain for the wave normal $\bp$, projected onto $\bu^M$, is $\sum_m\chi_m/[\omega^2-k^2(V^\mathrm{c}_m)^2]$, which has the same form as that of the reference medium in \cref{eq:G0}. In the notation of \cref{eq:modeterm}, wave $m$ then has $A=\chi_m/(V^\mathrm{c}_m)^2$, so a slow wave weighs more in the spectral function than its share suggests. After the average over the grain orientations, \cref{eq:partition} gives the partition of wave $m$ as
\begin{equation}\label{eq:raypartition}
 \eta^\mathrm{c}_m=\frac{\big\langle\chi_m/(V^\mathrm{c}_m)^2\big\rangle}{\sum_{n=1}^3\big\langle\chi_{n}/(V^\mathrm{c}_{n})^2\big\rangle},
\end{equation}
which is normalized to the sum over the three waves, like the partitions $\eta_i$ of the roots, so that the two can be compared directly in Table~\ref{tab:partition}. The third number is the relative r.m.s.\ spread $\zeta_m$ of the slowness about the mean. A wider band stays in phase over a shorter distance, so the spread sets how fast the contribution loses coherence and thus how strongly it is attenuated.

\begin{figure*}[!t]
 \centering
 \includegraphics[width=0.8\textwidth]{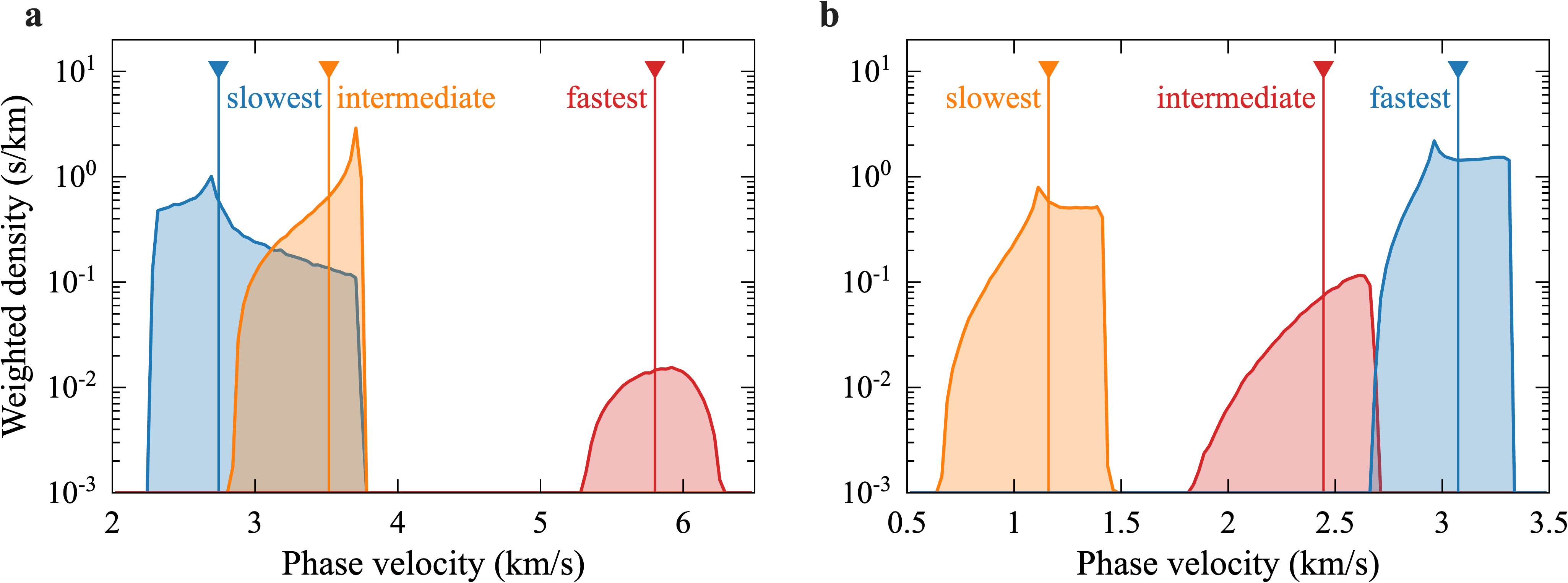}
 \caption{Grain velocity estimate: distributions of the phase velocities of the three crystal waves over the grain orientations, each weighted by $\chi_m$, \cref{eq:rayweight}. (a)~Transverse excitation in iron. (b)~Longitudinal excitation in SrMgSn. Triangles and vertical lines mark the grain velocities $\bar V_m$ of \cref{eq:rayvel}. Each wave is colored as the root it is compared with in Section~\ref{sec:results}: blue for root~1, orange for root~2 and red for root~3.}
 \label{fig:ray}
\end{figure*}

The estimate is evaluated for the two cases simulated in this paper, which are a transverse excitation in iron and a longitudinal excitation in SrMgSn. The averages are taken numerically over $4\times10^5$ orientations, with the wave normal $\bp$ drawn uniformly on the sphere in the crystal frame and the polarization $\bu^M$ taken along $\bp$ for a longitudinal excitation or drawn uniformly in the plane normal to $\bp$ for a transverse one. Figure~\ref{fig:ray} shows the weighted distributions of the phase velocities of the three crystal waves, and Table~\ref{tab:ray} lists the share, the grain velocity, the spread and the partition of each wave.

\begin{table*}[!t]
 \centering
 \caption{Grain velocity estimate. For each crystal wave, the table lists its share $\langle\chi_m\rangle$ of the excitation, the grain velocity $\bar V_m$ obtained from its mean slowness, the relative spread $\zeta_m$ of its slowness and its partition $\eta^\mathrm{c}_m$.}
 \label{tab:ray}
 \small
 \begin{tabular*}{\textwidth}{@{\extracolsep{\fill}}llrrrr}
 \toprule
 Material, wave & Crystal wave & $\langle\chi_m\rangle$ & $\bar V_m$ (m/s) & $\zeta_m$ (\%) & $\eta^\mathrm{c}_m$ \\
 \midrule
 Iron, T & Slowest & 0.49 & 2744 & 11.7 & 0.62 \\
 & Intermediate & 0.50 & 3517 & 6.2 & 0.38 \\
 & Fastest & 0.010 & 5802 & 3.8 & 0.003 \\
 \addlinespace
 SrMgSn, L & Slowest & 0.24 & 1161 & 14.1 & 0.69 \\
 & Intermediate & 0.038 & 2445 & 7.5 & 0.024 \\
 & Fastest & 0.72 & 3075 & 4.7 & 0.29 \\
 \bottomrule
 \end{tabular*}
\end{table*}

In iron, the slowest, the intermediate and the fastest waves are the slow transverse, the fast transverse and the longitudinal waves of the crystal. They have shares of 0.49, 0.50 and 0.010 of a transverse excitation, and partitions of 0.62, 0.38 and 0.003. The two transverse waves carry nearly equal shares, but the slowest has the larger partition because the spectral weight grows as the inverse square of the velocity.

In SrMgSn, the three waves cannot be identified as longitudinal or transverse waves because $c_{44}>c_{11}$, which makes the polarization of each wave change with direction. For example, the slowest wave is longitudinally polarized near the cube axes (1420~m/s along them), although it is transverse in most directions. The slowest, the intermediate and the fastest waves have shares of 0.24, 0.038 and 0.72 of a longitudinal excitation, and partitions of 0.69, 0.024 and 0.29. The share of the slowest wave comes mainly from directions near the cube axes, and its partition is again the largest.

In both materials, the slower of the two waves that carry a significant share of the excitation has the wider spread of slowness, by a factor of two in iron and three in SrMgSn. The estimate thus indicates that this slower wave loses its coherence over the shorter distance and therefore has the larger attenuation.

The estimate describes the formation of the coherent field in one layer of grains only and does not take into account refraction, the conditions at the grain boundaries or the scattering that follows. It therefore gives neither the attenuation coefficients nor the frequency at which the second mode appears. The estimate is compared with the theory and the simulations in Section~\ref{sec:results}.

\section{Results and discussion}\label{sec:results}

This section compares the theory, the simulations and the grain velocity estimate for transverse waves in iron (Section~\ref{sec:results_T}) and longitudinal waves in SrMgSn (Section~\ref{sec:results_L}). It then discusses what decides the observability of the modes (Section~\ref{sec:partition}), the relation to causality (Section~\ref{sec:causality}), the limits of the study (Section~\ref{sec:limitations}) and the generality of the results (Section~\ref{sec:generality}).

\subsection{Transverse waves in iron}\label{sec:results_T}

\begin{figure*}[!t]
 \centering
 \includegraphics[width=0.8\textwidth]{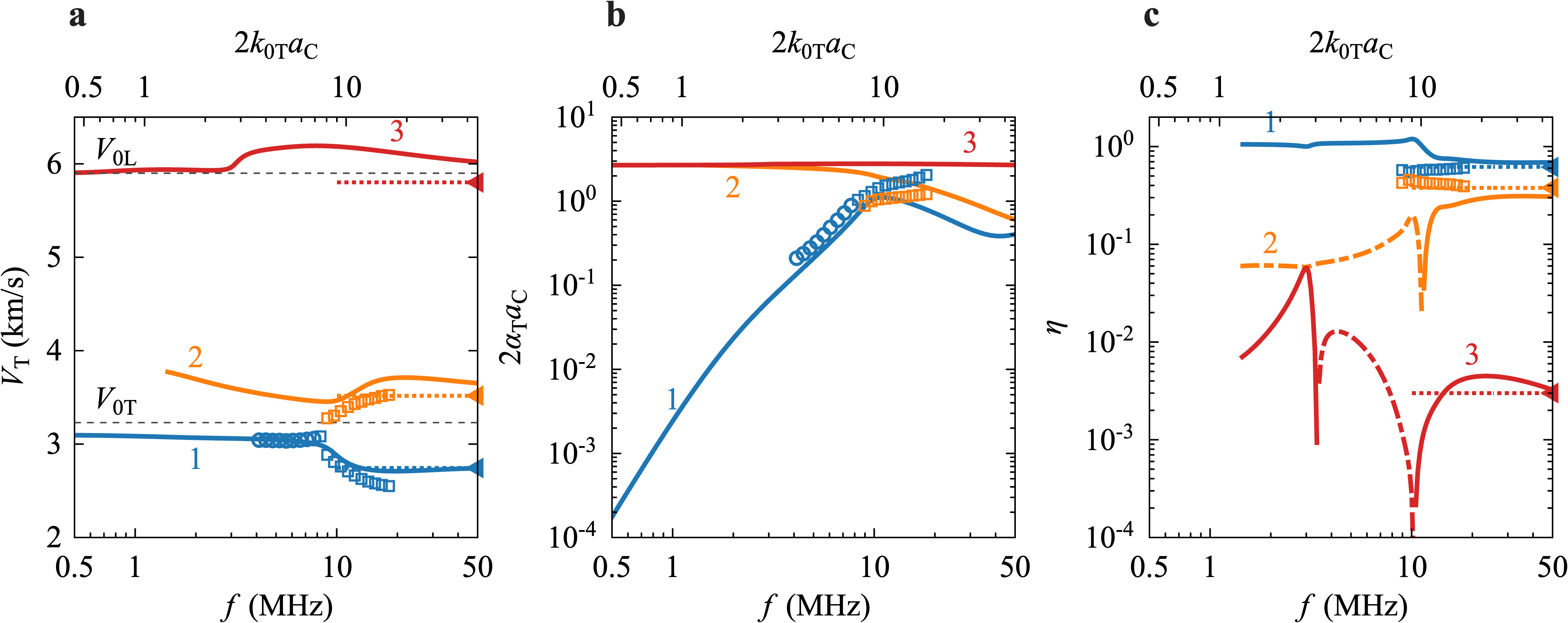}
 \caption{Transverse wave propagation in iron: theory, simulation and grain velocity estimate. (a)~Phase velocity $V_\mathrm{T}$ of the three roots of the dispersion equation, \cref{eq:dispersion}, numbered 1--3; horizontal dashed lines mark the Voigt transverse ($\vT$) and longitudinal ($\vL$) velocities, and dotted lines ending in triangles mark the grain velocities $\bar V_m$ of the three crystal waves (Table~\ref{tab:ray}) over the geometric regime, colored as the roots they are compared with. (b)~Attenuation coefficient normalized with the correlation length, $2\alpha_\mathrm{T}\aC$. (c)~Amplitude partition $\eta_i$ of each root, \cref{eq:partition}, normalized so that the three add to unity and shown where all three roots are tracked; dash-dotted parts of the curves show $-\eta_i$ where it is negative, and dotted lines ending in triangles mark the partitions $\eta^\mathrm{c}_m$ of the grain velocity estimate. Symbols in (a) and (b) are the values extracted from the simulated spectra of Fig.~\ref{fig:processing_T} by the Lorentzian fit, and those in (c) are the partitions $\eta^\mathrm{s}_i$ of the simulated modes, \cref{eq:simpartition}, where both modes are resolved. All symbols are colored as the corresponding root, with circles for the 8~MHz and squares for the 16~MHz simulation. Upper axes: normalized frequency $2\kT\aC$, with $\kT=2\pi f/\vT$.}
 \label{fig:result_T}
\end{figure*}

Figure~\ref{fig:result_T} compares theory, simulation and the grain velocity estimate for transverse waves in iron. The curves are the three roots of the dispersion equation, evaluated with the full FOSA mass operator, the symbols are the modes measured from the simulated spectra of Section~\ref{sec:processing}, and the dotted lines are the grain velocity estimate of Section~\ref{sec:ray}.

Theory, simulation and the grain velocity estimate agree closely in phase velocity, as shown in Fig.~\ref{fig:result_T}(a). Below the onset of the geometric regime ($2\kT\aC\simeq9$), a single mode is measured, and it follows root~1. Above the onset, a second mode appears close to where the theory predicts the attenuation of root~2 to fall to the level of root~1, and the two modes follow roots~1 and 2 to within about 5\% up to the highest simulated frequency ($2\kT\aC\simeq19$). The theory thus predicts both the velocities and the frequency at which the second mode appears. The three roots remain distinct in the complex wavenumber plane at all frequencies, as their velocities never cross. The grain velocity estimate makes no use of the scattering theory but agrees with the simulations (and the theory) comparably well. At 16~MHz (Table~\ref{tab:partition}), the grain velocity of the intermediate wave is within 1\% of the velocity of mode~2, compared with 5\% for root~2 of the theory. That of the slowest wave is 7\% above the velocity of mode~1 (at the same level as root~1), and this gap widens with frequency because mode~1 is still slowing down at the highest simulated frequency. The third root has no counterpart in the simulation at any frequency. A weak ridge near its velocity is observed in the off-polarization component $u_z$ in Appendix~\ref{sec:offpol}, but it is a residual of the ensemble average instead of a coherent mode.

The attenuation agrees less well, as shown in Fig.~\ref{fig:result_T}(b). Where a single mode propagates, the measured attenuation follows root~1 closely, and the attenuation of root~1 lies orders of magnitude below that of the other two roots. In the geometric regime, roots~1 and 2 have attenuation coefficients of the same order in both theory and simulation, whereas the attenuation of root~3 is 1.4 to 5 times that of root~2 and 2.6 to 7 times that of root~1 above 10~MHz. The measured attenuations depart from the predictions in opposite directions for the two modes. The attenuation of mode~1 exceeds the prediction by a factor that rises from 1.3 at 10~MHz to 3 at 19.5~MHz, whereas that of mode~2 is lower and rises from 0.5 to 0.9 of the prediction over the same range. The theory therefore reverses the order of the two attenuations because it predicts root~1 to be the less attenuated, whereas the simulations find mode~1 the more attenuated (Table~\ref{tab:partition}). Moreover, the theory predicts both attenuations to fall with frequency in this regime, while the measured attenuations of both modes rise slowly. The finite record length cannot account for this because both modes decay well within it (Section~\ref{sec:processing}). The discrepancy therefore lies in the theory, which is second order in the fluctuations and is used here at $2\alpha\aC\simeq1$ and above, where scattering is no longer weak. The theory thus predicts the velocities better than the attenuations, and attenuation in the geometric regime is the more demanding test of the approximation. The grain velocity estimate cannot predict the attenuation coefficients, but it indicates which of the two modes is the more attenuated. The slowest wave has the wider spread of slowness (11.7\% against 6.2\%) and should therefore lose its coherence over the shorter distance (Section~\ref{sec:ray}). The estimate thus expects mode~1 to be the more attenuated, which is the order found in the simulations and the opposite of that predicted by the theory.

Panel (c) shows the amplitude partition of each root, obtained from the residues of Section~\ref{sec:spectral}, together with the partitions of the simulated modes, \cref{eq:simpartition}, and of the grain velocity estimate. At low frequency, root~1 carries essentially all of the weight ($\eta_1\simeq1$), and the partition of root~2 is negative. A negative partition does not violate the positivity of the spectral function, which is a property of the sum. It means that root~2 does not form a peak of its own in this range but makes the line of root~1 asymmetric. As the geometric regime is approached, the partition of root~1 rises to a maximum near $2\kT\aC\simeq9$ and then falls, while that of root~2 passes through zero at $2\kT\aC=10.2$ (about 11.3~MHz) and becomes positive. The sign change of $\eta_2$ lies close to the onset at which the second ridge appears in the simulation (near $2\kT\aC\simeq9$). It offers a theoretical definition of the onset because root~2 can form a ridge of its own only where its partition is positive. In the geometric regime, roots~1 and 2 have partitions of about 0.70 and 0.30. The simulated modes carry 0.59 and 0.41 at 16~MHz and nearly the same from 10 to 18~MHz (Table~\ref{tab:partition}), and the slowest and the intermediate waves of the grain velocity estimate carry 0.62 and 0.38, within 0.03 of the simulations. The theory is further away mainly because the residues of roots~1 and 2 are strongly complex, and it gives 0.61 and 0.38 when its partitions are formed from $|R_i|$, as for the simulated modes. Root~3 has a partition of about 0.004 in the geometric regime, close to the 0.003 of the fastest wave in the estimate. Its partition is also below 0.01 at low frequency, except near 3~MHz, where it reaches about 0.06 and changes sign because root~3 nearly coincides with another heavily damped root.

The three roots explain the physical origins of these observations. At low frequency ($2\kT\aC<9$), root~1 is the effective wave of the classical description. It propagates marginally slower than $\vT$, near the quasi-static velocity limit of the FOSA (Section~\ref{sec:fosa}), and it is the only weakly attenuated root. This is why conventional ultrasonic measurements observe only one transverse wave. In the geometric regime beyond $2\kT\aC\simeq9$, the wavelength is shorter than the grains and the wave samples the local wave speed of each grain. The scattering factors of iron, evaluated from the single-crystal constants following \citet{Rokhlin2015}, show that a transverse wave scatters mainly into its own channel ($Q_{T\to T}=0.018$ against $Q_{T\to L}=0.003$). Iron therefore follows the first case of Section~\ref{sec:ffa}, in which root~2 is the transverse-channel root and pairs with root~1. \Cref{eq:geompair} places the pair on either side of $\vT$ by $\sqrt{Q_{T\to T}}=13.5\%$ and predicts 2.82 and 3.70~km/s, which are close to the 2.7 and 3.7~km/s approached by the FOSA. Within the theory, roots~1 and~2 are thus the slow and the fast parts of the distribution of local transverse wave speeds. Both transverse polarizations have the same velocity $\vT$ in the isotropic reference medium, and the two roots are split by the fluctuation of that velocity from grain to grain.

The grain velocity estimate confirms these physical origins and identifies the two parts with the slowest and the intermediate waves of the individual grains (the slow and the fast transverse waves), whose grain velocities are 2.74 and 3.52~km/s (Section~\ref{sec:ray}). The theory and the estimate are consistent. The theory represents the two transverse waves of the grains by a single reference velocity with a fluctuation of $\sqrt{Q_{T\to T}}=13.5\%$, and the phase velocities of the two waves in the grains, pooled into one distribution with their weights $\chi_m$, have a comparable r.m.s.\ spread of 14.9\%. Root~3 stays near the longitudinal channel pole and describes transverse motion converted into longitudinal motion, so its attenuation is set by the grain size instead of the frequency (Section~\ref{sec:ffa}). It corresponds to the fastest (longitudinal) wave of the grains, whose share of a transverse excitation is only 0.01.

Roots~1 and 2 thus have attenuation coefficients of the same order and comparable partitions in the geometric regime, so both survive over comparable distances and appear in the spectra as modes~1 and~2. Root~3 combines the largest attenuation with a partition below 0.01 and is never observed, as Section~\ref{sec:partition} quantifies.

\subsection{Longitudinal waves in SrMgSn}\label{sec:results_L}

\begin{figure*}[!t]
 \centering
 \includegraphics[width=0.8\textwidth]{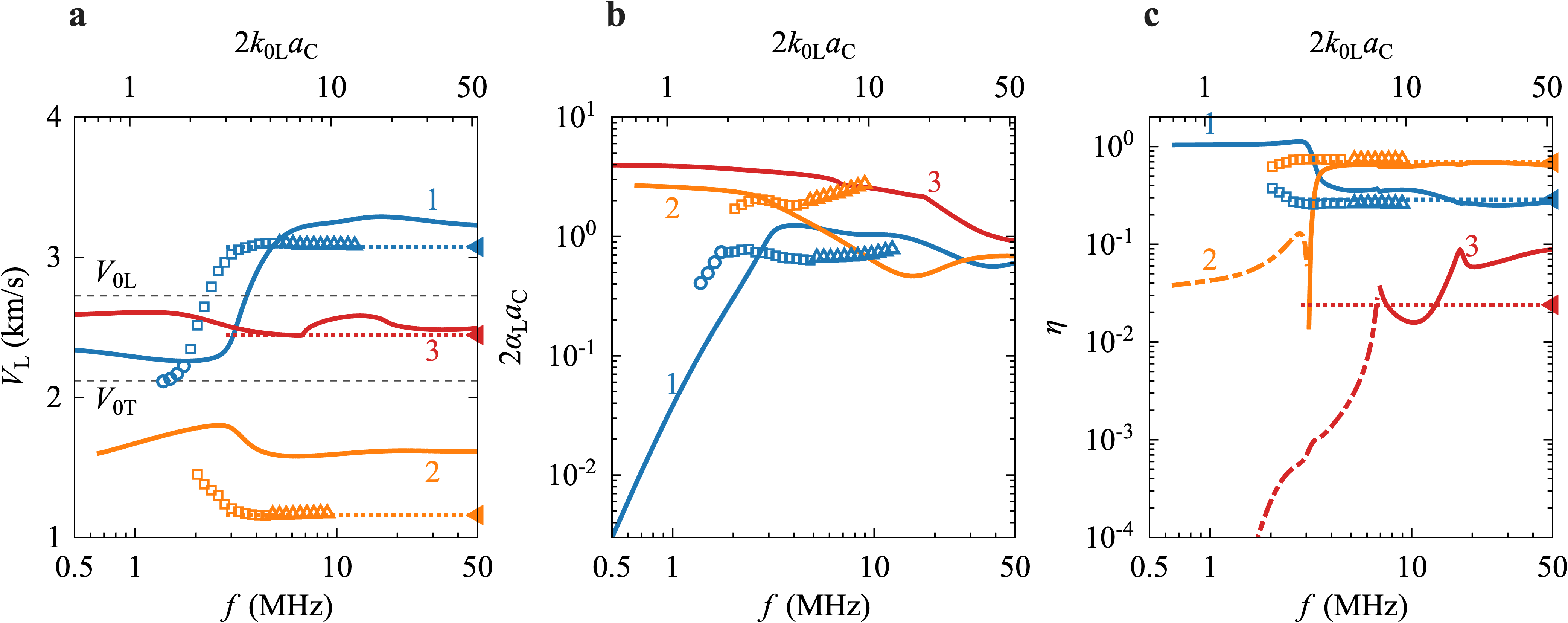}
 \caption{Longitudinal wave propagation in SrMgSn: theory, simulation and grain velocity estimate. Same as Fig.~\ref{fig:result_T}, for the strongly anisotropic polycrystal SrMgSn simulated at 2 (circles), 4 (squares) and 8~MHz (triangles) center frequency. Upper axes: normalized frequency $2\kL\aC$, with $\kL=2\pi f/\vL$.}
 \label{fig:result_L}
\end{figure*}

Figure~\ref{fig:result_L} compares theory, simulation and the grain velocity estimate for longitudinal waves in SrMgSn.

The same phenomenon appears in phase velocity in Fig.~\ref{fig:result_L}(a), with a single mode at low frequency and two modes above the onset of the geometric regime. The onset occurs at $2\kL\aC\simeq3$ (where $\lambda_\mathrm{0L}\simeq4\aC$), which is earlier in normalized frequency than in iron because the scattering is stronger. The agreement between theory and simulation is looser than in iron because the FOSA is pushed well outside the weak-scattering conditions for which it was derived. Mode~1 follows root~1 to within about 10\% below the onset and 6\% above it, although its velocity rises through the transition somewhat earlier (near $2\kL\aC\simeq2$--3), and mode~2 lies 25 to 35\% below root~2. The grain velocity estimate accounts for the measured velocities more closely than the theory does. The grain velocities of the fastest and the slowest waves (3.08 and 1.16~km/s) are within about 1\% of the velocities of modes~1 and~2 simulated at 8~MHz (3.09 and 1.17~km/s), whereas roots~1 and~2 are 5\% and 35\% above them (Table~\ref{tab:partition}). As in iron, the third root has no counterpart in the simulation. Its velocity crosses that of root~1 near 3~MHz, but the two roots do not coincide there because their attenuations differ by a factor of three.

The attenuation in Fig.~\ref{fig:result_L}(b) agrees even less well for the same reason. At low frequency, root~1 is the only weakly attenuated root. In the geometric regime, roots~1 and 2 attain comparable attenuation, whereas root~3 attenuates two to five times faster than the others in the simulated range above 7~MHz. The measured attenuation of mode~1 lies below that of root~1 by a factor of about 1.5, and that of mode~2 lies above that of root~2 by a factor of up to about 5. As in iron, the simulation finds the fast mode the less attenuated, whereas the theory predicts the slow root to be the less attenuated above $2\kL\aC\simeq6$ (Table~\ref{tab:partition}). The grain velocity estimate again indicates the order found in the simulations, because the slowest wave of the grains has the wider spread of slowness (14.1\% against 4.7\%) and the slow mode is thus expected to be the more attenuated.

Panel (c) shows the same pattern of amplitude partitions as in iron. The partition of root~2 is negative at low frequency and passes through zero at $2\kL\aC=3.25$ (about 3.05~MHz), which is again close to the onset at which the second mode appears in the simulation. In the geometric regime, roots~1 and~2 have partitions of about 0.35 and 0.65. The simulated modes carry 0.26 and 0.74 at 8~MHz (Table~\ref{tab:partition}), and the fastest and the slowest waves of the grain velocity estimate carry 0.29 and 0.69, within 0.05 of the simulations. Unlike in iron, the theory stays further from the simulations than the estimate, even when its partitions are formed from $|R_i|$. The partition of root~3 is negligible at low frequency and grows to 0.07 at 30~MHz, whereas the intermediate wave of the estimate carries 0.024.

The roots have the same origins as in iron, but SrMgSn follows the second case of Section~\ref{sec:ffa}, in which root~1 pairs with the root of the other channel. The transverse channel is strongly coupled to the longitudinal wave because its scattering factor is large ($Q_{L\to T}=0.044$ against $Q_{L\to L}=0.025$, both independent of frequency) and the two reference velocities are close ($\vL/\vT=1.29$). At low frequency, root~1 is the effective wave, which lies below $\vL$ near the quasi-static velocity limit of the FOSA and is the only one observed. In the geometric regime, it pairs with the transverse-channel root (root~2), and their strong coupling pushes root~1 above $\vL$ to 3.2~km/s and root~2 below $\vT$ to about 1.6~km/s. Root~3 stays near the longitudinal channel pole at about 2.5~km/s. Within the theory, root~2 thus carries the part of the longitudinal excitation that travels as transverse motion between scattering events. The grain velocity estimate gives the physical origin of the two modes at the grain scale. Mode~1 is the fastest wave of the grains, and mode~2 is the slowest wave, which is transverse in most directions but is longitudinally polarized near the cube axes, where a longitudinal excitation drives it (Section~\ref{sec:ray}). The intermediate wave has a share of only 0.04 and lies near root~3.

Roots~1 and 2 thus appear in the spectra as modes~1 and~2, whereas root~3 is not observed (Section~\ref{sec:partition}). Theory and simulation therefore agree on the structure of the coherent field, which has a single mode at low frequency, a second mode from the onset of the geometric regime and no third mode. This robustness at $A^U=17.67$ (an anisotropy an order of magnitude beyond that of iron) is surprising for a second-order approximation, which loses accuracy for strongly anisotropic polycrystals even at low frequency \citep{Huang2022a} and whose quasi-static velocity for SrMgSn is already 10\% above the self-consistent estimate (Section~\ref{sec:fosa}). It indicates that the multimodal structure is not an artifact of weak scattering.

In both materials, the agreement between the grain velocity estimate and the simulations is striking. The estimate has no adjustable parameter and uses only the single-crystal elastic constants, but three of the four observed modes lie within 1\% of it in velocity, and all four within 0.05 in partition, in two materials whose anisotropy differs by an order of magnitude. The agreement does not hinge on the kind of average because the weighted mean of the phase velocities differs from $\bar V_m$ by less than 2\%, but it requires the weights $\chi_m$, without which the slowest wave in SrMgSn would be 12\% below mode~2. We find such a close agreement surprising for an estimate confined to one layer of grains. The estimate already shows that the coherent modes of the geometric regime are formed from the plane waves that the excitation launches in the grains. Their existence now rests on three independent descriptions, which are the scattering theory, the grain-scale simulations and the grain velocity estimate.

\subsection{Observability of the modes}\label{sec:partition}

In our previous work, the number of observable modes was attributed to the attenuation hierarchy of the roots \citep{Huang2026prl}. The spectral-function analysis of Section~\ref{sec:spectral} shows that attenuation is only one of two factors. Table~\ref{tab:partition} gives both factors, the attenuation and the amplitude partition, for each root, together with their combined effect, the ridge height of \cref{eq:height}. 

\begin{table*}[!t]
 \centering
 \caption{Velocities, attenuations and amplitude partitions in the geometric regime from the FOSA, the simulations and the grain velocity estimate. At representative frequencies, the table lists for each root of the FOSA the phase velocity $V_i$, the normalized attenuation $2\alpha_i\aC$, the amplitude partition $\eta_i$ normalized to the sum over the three roots, and the ridge height $H_i=A_i/(2k_i\alpha_i)$ of the spectral function relative to root~1. The simulation gives the phase velocity, the normalized attenuation and the partition $\eta^\mathrm{s}_i$, \cref{eq:simpartition}, of the modes observed at 16~MHz in iron and 8~MHz in SrMgSn, and dashes mark the other frequencies and root~3, which is never observed. The grain velocity estimate gives the grain velocity $\bar V_m$ and the partition $\eta^\mathrm{c}_m$, which do not depend on frequency.}
 \label{tab:partition}
 \small\setlength{\tabcolsep}{4pt}
 \begin{tabular*}{\textwidth}{@{\extracolsep{\fill}}lrrrrrrrrrrrr}
 \toprule
 & & & & \multicolumn{4}{c}{FOSA} & \multicolumn{3}{c}{Simulation} & \multicolumn{2}{c}{Grain velocity estimate} \\
 \cmidrule(lr){5-8} \cmidrule(lr){9-11} \cmidrule(l){12-13}
 Material, wave & \shortstack[r]{$f$\\(MHz)} & $2k_{0M}\aC$ & Root & $V_i$ (m/s) & $2\alpha_i\aC$ & $\eta_i$ & $H_i/H_1$ & $V_i$ (m/s) & $2\alpha_i\aC$ & $\eta^\mathrm{s}_i$ & $\bar V_m$ (m/s) & $\eta^\mathrm{c}_m$ \\
 \midrule
 Iron, T & 16 & 14.4 & 1 & 2716 & 0.89 & 0.75 & 1 & 2573 & 1.83 & 0.59 & 2744 & 0.62 \\
  & & & 2 & 3679 & 1.59 & 0.25 & 0.26 & 3503 & 1.17 & 0.41 & 3517 & 0.38 \\
  & & & 3 & 6142 & 2.78 & 0.004 & 0.004 & -- & -- & -- & 5802 & 0.003 \\
 \addlinespace
 Iron, T & 30 & 27.0 & 1 & 2723 & 0.47 & 0.69 & 1 & -- & -- & -- & 2744 & 0.62 \\
  & & & 2 & 3697 & 0.97 & 0.31 & 0.30 & -- & -- & -- & 3517 & 0.38 \\
  & & & 3 & 6070 & 2.74 & 0.004 & 0.002 & -- & -- & -- & 5802 & 0.003 \\
 \addlinespace
 SrMgSn, L & 8 & 8.5 & 1 & 3235 & 1.05 & 0.36 & 1 & 3089 & 0.68 & 0.26 & 3075 & 0.29 \\
  & & & 2 & 1585 & 0.78 & 0.62 & 1.14 & 1172 & 2.57 & 0.74 & 1161 & 0.69 \\
  & & & 3 & 2526 & 2.64 & 0.021 & 0.018 & -- & -- & -- & 2445 & 0.024 \\
 \addlinespace
 SrMgSn, L & 30 & 31.9 & 1 & 3256 & 0.61 & 0.25 & 1 & -- & -- & -- & 3075 & 0.29 \\
  & & & 2 & 1617 & 0.65 & 0.68 & 1.25 & -- & -- & -- & 1161 & 0.69 \\
  & & & 3 & 2484 & 1.21 & 0.070 & 0.11 & -- & -- & -- & 2445 & 0.024 \\
 \bottomrule
 \end{tabular*}
\end{table*}

Roots~1 and 2 are observable because both factors favor them. Their partitions are comparable in the geometric regime (roughly 0.70 and 0.30 in iron and 0.35 and 0.65 in SrMgSn), and their attenuations are of the same order, so their ridges are of comparable height. The simulations agree, with two ridges of comparable height and partitions of 0.59 and 0.41 in iron and 0.26 and 0.74 in SrMgSn (Table~\ref{tab:partition}).

Root~3 is suppressed by both factors. In iron, its partition is 0.004 and it attenuates 1.7 to 6 times faster than the others, and the two factors multiply in \cref{eq:height} to place its ridge between two and three orders of magnitude below that of root~1. In SrMgSn at 8~MHz, it attenuates 2.5 to 3.4 times faster than the others and its partition is 0.02, so its ridge is nearly two orders of magnitude below them. The spectra of Figs.~\ref{fig:processing_T} and \ref{fig:processing_L} span two decades, and the residual fluctuation of the twenty-realization average, which the off-polarization components quantify (Appendix~\ref{sec:offpol}), lies within the lowest decade, so a ridge two or more decades below the strongest cannot be seen. This is why root~3 is never visible in the numerical experiments, and this explanation is stronger than the attenuation argument because the attenuation alone would not exclude root~3 in SrMgSn. Its partition in SrMgSn grows with frequency and reaches 0.07 at 30~MHz, where its ridge is only one decade below that of root~1, so whether root~3 becomes observable at higher frequency in strongly anisotropic materials is left for future work.

The amplitude partition also indicates when a second mode should be expected in other materials. Two transverse modes should be general, because a transverse excitation always divides between the two transverse waves of an anisotropic crystal. A second longitudinal mode like that of SrMgSn requires strong anisotropy. In SrMgSn, the slowest wave of the grains is the most longitudinally polarized of the three waves in 18\% of the directions and has a share of 0.24 of a longitudinal excitation, whereas in iron the fastest wave has a share of 0.98, with a polarization that deviates from the wave normal by at most 12$^\circ$.

The theory reflects this difference through the channel whose root pairs with root~1. In common materials such as iron, aluminum and titanium, a longitudinal wave scatters mainly into longitudinal waves at high frequency, after a transition marked by a hump of the attenuation at the beginning of the stochastic regime \citep{Hirsekorn1982,Stanke1984,Yang2012}, whereas the attenuation of mode~1 in SrMgSn shows no such hump (Fig.~\ref{fig:result_L}(b)). For longitudinal waves in iron, the transverse channel is far in velocity ($\vL/\vT=1.83$) and weakly coupled, and \cref{eq:geomcubic} gives the first case of Section~\ref{sec:ffa}, with a pair of roots on either side of $\vL$ by $\sqrt{Q_{L\to L}}=4.3\%$. This pair represents the slow and the fast parts of a single band of phase velocities in the grains, and whether it can be resolved as two modes remains to be examined.

\subsection{Causality and the two-mode description}\label{sec:causality}

The velocity dispersion of the roots bears on the causality argument of \citet{Calvet2012}. For a single causal wave with positive attenuation, the Kramers--Kronig relations require the phase velocity to be lower in the low-frequency Rayleigh limit than in the high-frequency geometric limit \citep{WeaverPao1981,ODonnell1981,Beltzer1986}. In iron, root~1 violates this requirement because its velocity falls from 3.1~km/s at low frequency to 2.7~km/s in the geometric regime, as shown in Fig.~\ref{fig:result_T}(a). It is therefore not a causal wave on its own and cannot describe the coherent field in the geometric regime. The velocity increase demanded by causality is carried by root~2, which propagates at 3.7~km/s. The sum rule of \cref{eq:sumruleAB} ties the two together because root~2 takes over part of the spectral weight lost by root~1 as the geometric regime is approached. The observation of both modes in iron thus confirms that a single-wave description of the coherent field in the geometric regime is not merely inaccurate but forbidden, and it shows that the root restoring causality is observed. The roles are exchanged in SrMgSn, where root~1 speeds up through the transition, as a single causal wave may, whereas root~2 slows down. Causality alone does not require the second root there, but the simulations still show it as the mode that carries most of the spectral weight in the geometric regime (Section~\ref{sec:partition}).

\subsection{Limits of the theory, the simulation and the estimate}\label{sec:limitations}

Several limitations should be stated. On the theory side, the FOSA is second order in the fluctuations, and its attenuations depart from the simulations at $2\alpha\aC\simeq1$ and above (Section~\ref{sec:results_T}). Recurrent scattering, which the present theory excludes and which \citet{Roy2025} found to matter most for transverse waves, may contribute to this departure. In SrMgSn, the theory also overestimates the velocity of the slow mode by 25 to 35\% (Section~\ref{sec:results_L}). The grain velocity estimate suggests that this is because the single transverse velocity of the isotropic reference medium (2.1~km/s) stands for the slowest and the intermediate waves of the SrMgSn crystal, which lie at about 1.2 and 2.4~km/s. The theory is therefore qualitatively but not quantitatively correct at $A^U=17.67$. Finally, root~3 lies among the additional heavily damped roots of the multi-term correlation function below about 7~MHz (Section~\ref{sec:ffa}) and nearly coincides with some of them, where it is followed continuously in frequency. This does not affect the conclusions because root~3 cannot be observed.

On the simulation side, the finite record length and the interference between modes limit the accuracy of the fitted attenuations as described in Section~\ref{sec:processing}, and the bias from the record length could be removed where needed by fitting the exact line shape \citep{Huang2026jasa}. The twenty-realization average leaves a residual incoherent field of up to about 1\% of the coherent peak, which sets the noise floor of the spectra, and the cross-section average over a $12\times12$~mm$^2$ face (about 24 grains across) contributes to the same residual.

The grain velocity estimate is confined to one layer of grains. It predicts the velocities and the partitions of the modes, but neither their attenuation coefficients nor the frequency at which the second mode appears (Section~\ref{sec:ray}).

\subsection{Generality}\label{sec:generality}

In the mechanism identified here, each scattering channel contributes a coherent wave besides the effective wave, and one of these waves joins the effective wave once the wavelength resolves the microstructure. The derivation of Section~\ref{sec:ffa} shows that the number of roots is one more than the number of distinct channel velocities for a single-exponential correlation function, and the analysis of Section~\ref{sec:spectral} shows that the spectral weight is conserved and redistributed among them. The Dyson equation and its dispersion relation govern coherent propagation in all random media \citep{Frisch1968,Sheng2006}. Comparable multimodality should thus arise for electromagnetic waves in random dielectrics, whose single transverse channel gives two roots, as for scalar waves. It should also arise for high-frequency seismic waves in the heterogeneous Earth and other planetary bodies, whose longitudinal and transverse channels are the same as here. The same reasoning answers for polycrystals the question raised for particulate materials by \citet{Gower2019a}. Among the many mathematically admissible effective waves, the surviving ones are those whose attenuation permits propagation over many microstructural lengths and whose partition is not negligible. The result also bears on theories of high-frequency energy transport \citep{Weaver1990}.

\section{Conclusions}\label{sec:conclusions}

We have studied coherent elastic wave propagation in statistically isotropic polycrystals by theory, grain-scale simulation and a grain velocity estimate. We derived the dispersion equation of the first-order smoothing approximation, used the far-field approximation to make the structure of its roots explicit, and introduced a method that obtains the amplitude of every root from the residues of the mean Green's function, without fitting. We computed the coherent field of transverse waves in iron and of longitudinal waves in the strongly anisotropic SrMgSn by three-dimensional finite-element simulation of Voronoi polycrystals and resolved its modes in the frequency--wavenumber plane. We also estimated the modes of the geometric regime from the plane waves of the single crystal, without the scattering theory or the simulations.

The work leads to five general findings.
\begin{enumerate}
 \item The coherent field of a single-phase random polycrystal carries two modes instead of one once the wavelength becomes shorter than the grains. Two transverse modes were observed in iron and two longitudinal modes in SrMgSn, and they arise because the incident plane wave is extinguished within about one correlation length and the coherent field beyond is rebuilt by forward multiple scattering.
 \item The modes originate in the scattering channels. The scattering channel structure of the mass operator gives the dispersion equation exactly three roots, which correspond to the effective wave of the classical description and to one wave for each distinct scattering channel, and the effective wave is joined by one of the channel waves in the geometric regime.
 \item The modes have a direct physical meaning at the grain scale. They are formed from the plane waves that the excitation launches in the individual grains, and the grain velocity estimate built on these waves gives velocities within 1\% of both simulated modes in SrMgSn and of the fast mode in iron, and within 7\% of the slow mode in iron, without any adjustable parameter.
 \item The observability of a root is decided by its attenuation and its amplitude partition together, through the height of its ridge in the spectral function, \cref{eq:height}. Roots~1 and 2 have comparable attenuations and partitions in the geometric regime, and the partition of root~2 turns positive close to where the second mode appears, whereas root~3 combines a small partition with the largest attenuation and is never observed. The number of observable coherent modes is therefore set neither by the number of roots nor by attenuation alone, and the new residue method gives the partitions for any random medium and any approximation of the mass operator.
 \item The first-order smoothing approximation captures the structure of the coherent field in both materials, whose anisotropy differs by an order of magnitude, and the velocities in iron, but it does not accurately reproduce the attenuations in the geometric regime, including which mode is the less attenuated.
\end{enumerate}

The single effective wave of the classical description thus holds through the Rayleigh and stochastic regimes, which cover most practical ultrasonic measurements, but measurements of velocity and attenuation in the geometric regime must allow for two coherent modes. Two transverse modes should be general in anisotropic polycrystals, whereas a second longitudinal mode like that of SrMgSn requires strong anisotropy. Two questions remain open. The first is the attenuation of the two modes in the geometric regime, which calls for a theory beyond first-order smoothing. A grain-scale ray-tracing model would complement such a theory by showing how the modes are formed and attenuated as the waves pass from grain to grain, and why the grain velocity estimate agrees so closely with the simulations. The second is whether the third root becomes observable at higher frequency in strongly anisotropic materials.

\appendix
\setcounter{figure}{0}
\setcounter{table}{0}
\renewcommand{\thefigure}{A\arabic{figure}}
\section{Off-polarization components of the coherent field}\label{sec:offpol}

\begin{figure*}[!t]
 \centering
 \includegraphics[width=0.8\textwidth]{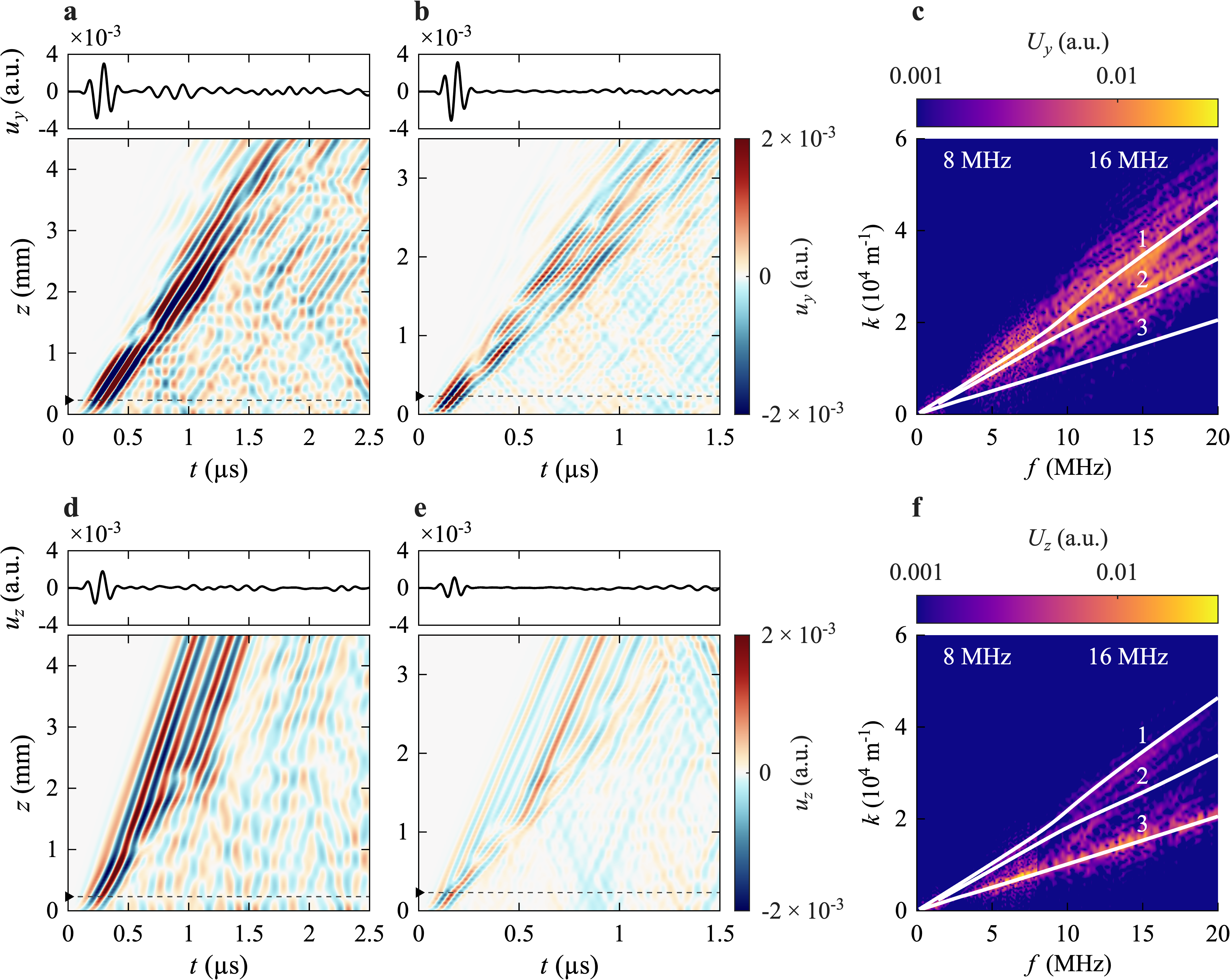}
 \caption{Off-polarization components of the coherent field for transverse excitation in iron. (a,~b)~Component $u_y$ at 8~MHz (a) and 16~MHz (b) center frequency, each averaged over the cross section and over 20 realizations and normalized to the peak of the excited component $u_x$ of the same simulation; the strip above each map is the waveform at $z=0.25~\mathrm{mm}\simeq\aC$. (c)~Magnitude of the two-dimensional Fourier transform of $u_y$, assembled from the 8~MHz simulation up to 10~MHz and from the 16~MHz simulation from 8~MHz upward, normalized to the maximum of the transform of $u_x$ and shown on a logarithmic color scale; the white curves are the three roots of the dispersion equation for transverse waves. (d--f)~The same as (a--c) for the component $u_z$. Color scales saturate at $\pm2\times10^{-3}$ in (a,~b,~d,~e).}
 \label{fig:offpol_T}
\end{figure*}

\begin{figure*}[!t]
 \centering
 \includegraphics[width=0.8\textwidth]{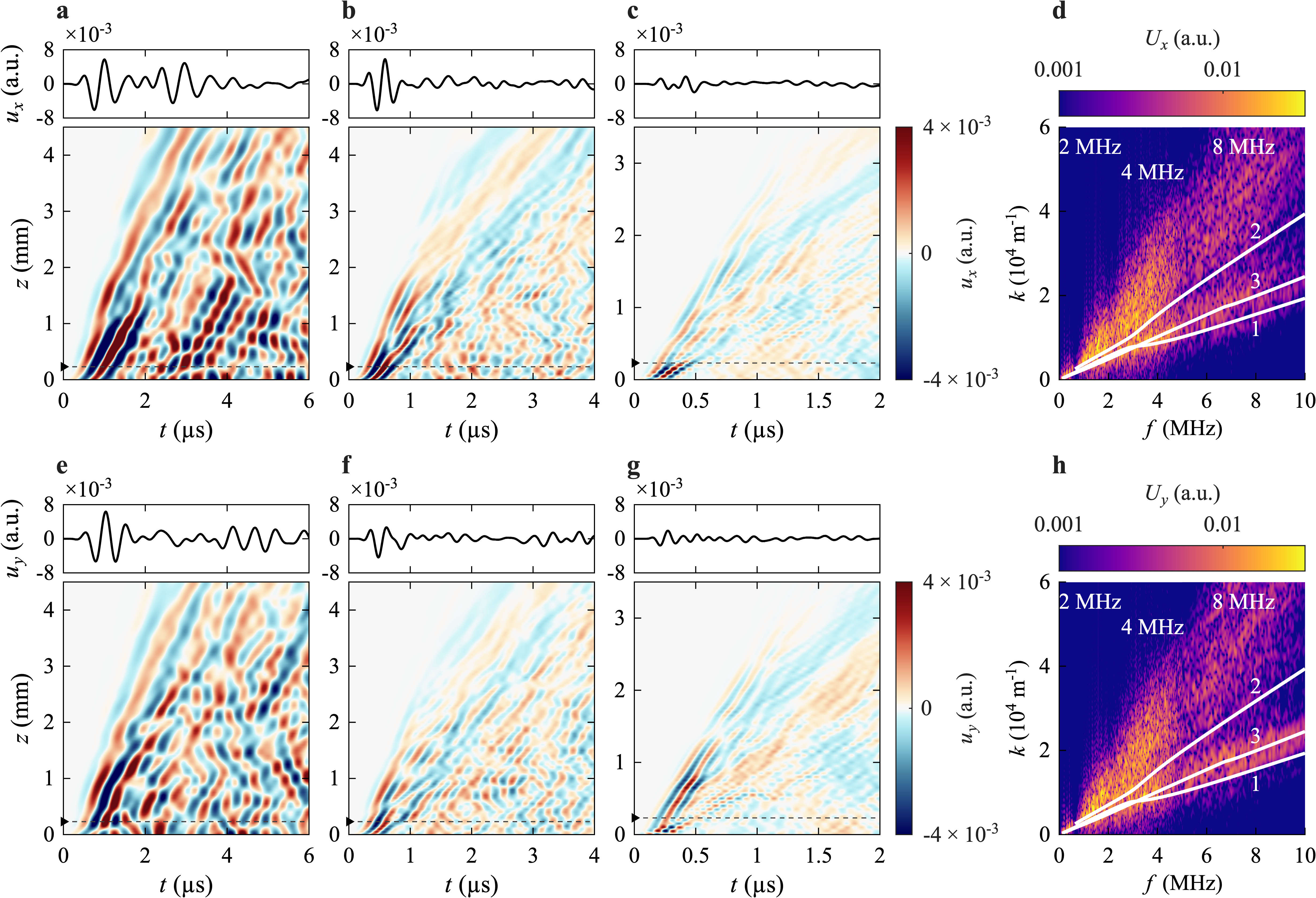}
 \caption{Off-polarization components of the coherent field for longitudinal excitation in SrMgSn. (a--c)~Component $u_x$ at 2~MHz (a), 4~MHz (b) and 8~MHz (c) center frequency, each averaged over the cross section and over 20 realizations and normalized to the peak of the excited component $u_z$ of the same simulation. (d)~Magnitude of the two-dimensional Fourier transform of $u_x$, assembled from the 2~MHz simulation up to 4~MHz, the 4~MHz simulation from 2 to 6~MHz and the 8~MHz simulation from 5~MHz upward, normalized to the maximum of the transform of $u_z$; the white curves are the three roots of the dispersion equation for longitudinal waves. (e--h)~The same as (a--d) for the component $u_y$. Color scales saturate at $\pm4\times10^{-3}$ in (a--c,~e--g).}
 \label{fig:offpol_L}
\end{figure*}

The components of displacement orthogonal to the excited polarization are two to three orders of magnitude smaller than the excited component and carry no coherent part, as Figs.~\ref{fig:offpol_T} and \ref{fig:offpol_L} show. For a statistically isotropic medium, the mean Green's function of \cref{eq:G0} has no coupling between the longitudinal and transverse polarizations. The mean field of a transverse excitation thus has no $u_z$ component, and it has no $u_y$ component by symmetry. Similarly, the mean field of a longitudinal excitation has no $u_x$ or $u_y$. The simulations therefore record in these components the residual of the incoherent field that twenty realizations have not averaged out, and its size measures the convergence of the average.

For the transverse excitation in iron shown in Fig.~\ref{fig:offpol_T}, the peak amplitudes of the averaged $u_y$ and $u_z$ are $7\times10^{-3}$ and $3\times10^{-3}$ of the peak of $u_x$ at 8~MHz, and $3\times10^{-3}$ and $1\times10^{-3}$ at 16~MHz, and their spectra peak at 2 to 4\% of the spectral peak of $u_x$. The maps (a,b) show that the residual in $u_y$ is an irregular pattern of streaks whose slopes span the transverse velocities, and its spectrum (c) is a diffuse band between about 2.5 and 3.5~km/s that overlaps the two transverse roots without following either. The residual in $u_z$ in maps (d,e) is different in character because it forms a narrow ridge in (f) at about 5.9~km/s (the Voigt longitudinal velocity) at all frequencies. It is a weak longitudinal disturbance launched where the applied transverse force meets the anisotropic surface grains and converts partly into longitudinal motion. Its sign varies with the orientation of the surface grains, so it averages toward zero and leaves the residual seen here after twenty realizations. Its velocity is close to that of root~3 because both are tied to the longitudinal velocity, but it is not root~3 because root~3 decays within about one correlation length ($2\alpha\aC\simeq2.7$), whereas the ridge persists over the whole record. The ridge is therefore not a coherent mode of the transverse excitation.

For the longitudinal excitation in SrMgSn shown in Fig.~\ref{fig:offpol_L}, the residuals are relatively larger, as expected for the stronger scattering. The peak amplitudes of $u_x$ and $u_y$ are between $4\times10^{-3}$ and $1\times10^{-2}$ of the peak of $u_z$, and their spectra peak at 5 to 10\% of the spectral peak of $u_z$. Both components show the same diffuse spectrum (d,h), whose energy is concentrated between about 1.0 and 1.5~km/s at 2 to 4~MHz, around the transverse-channel root (root~2), and spreads over the transverse velocities at 8~MHz. This is the expected signature of the incoherent transverse field generated by mode conversion into the transverse channel (the dominant one in SrMgSn), and it contains no ridge that the excited component does not already show. The off-polarization components thus confirm in both materials that the coherent modes reported in this paper are carried entirely by the excited polarization. They also show that the residual incoherent field lies one to two orders of magnitude below the coherent peak in the spectra, and this residual sets the floor against which the absence of root~3 is judged.

\bibliography{references}

@article{Biot1956a,
	title = {Theory of propagation of elastic waves in a fluid-saturated porous solid. {I}. {Low}-frequency range},
	volume = {28},
	doi = {10.1121/1.1908239},
	number = {2},
	journal = {The Journal of the Acoustical Society of America},
	author = {Biot, M. A.},
	year = {1956},
	pages = {168--178}
}

@article{Biot1956b,
	title = {Theory of propagation of elastic waves in a fluid-saturated porous solid. {II}. {Higher} frequency range},
	volume = {28},
	doi = {10.1121/1.1908241},
	number = {2},
	journal = {The Journal of the Acoustical Society of America},
	author = {Biot, M. A.},
	year = {1956},
	pages = {179--191}
}

@article{Karal1964,
author = {Karal, Frank C. and Keller, Joseph B.},
doi = {10.1063/1.1704145},
journal = {Journal of Mathematical Physics},
number = {4},
pages = {537--547},
publisher = {American Institute of Physics},
title = {{Elastic, electromagnetic, and other waves in a random medium}},
volume = {5},
year = {1964}
}

@incollection{Frisch1968,
author = {Frisch, Uriel},
booktitle = {Probabilistic Methods in Applied Mathematics},
editor = {Bharucha-Reid, A. T.},
pages = {75--198},
publisher = {Academic Press, New York},
title = {{Wave propagation in random media}},
year = {1968}
}

@book{Ishimaru1978,
author = {Ishimaru, Akira},
publisher = {Academic Press},
title = {{Wave Propagation and Scattering in Random Media}},
year = {1978},
address = {New York, NY},
doi = {10.1016/B978-0-12-374701-3.X5001-7}
}

@article{Plona1980,
	title = {Observation of a second bulk compressional wave in a porous medium at ultrasonic frequencies},
	volume = {36},
	issn = {00036951},
	doi = {10.1063/1.91445},
	number = {4},
	journal = {Applied Physics Letters},
	author = {Plona, Thomas J.},
	year = {1980},
	pages = {259--261}
}

@article{Stanke1984,
author = {Stanke, Fred E. and Kino, G. S.},
doi = {10.1121/1.390577},
journal = {The Journal of the Acoustical Society of America},
number = {3},
pages = {665--681},
title = {{A unified theory for elastic wave propagation in polycrystalline materials}},
volume = {75},
year = {1984}
}

@book{Rytov1989,
address = {New York},
author = {Rytov, Sergei M. and Kravtsov, Yurii A. and Tatarskii, Valeryan I.},
publisher = {Springer},
title = {{Principles of Statistical Radiophysics 4: Wave Propagation Through Random Media}},
year = {1989}
}

@article{Weaver1990,
author = {Weaver, R.L.},
doi = {10.1016/0022-5096(90)90021-U},
journal = {Journal of the Mechanics and Physics of Solids},
number = {1},
pages = {55--86},
title = {{Diffusivity of ultrasound in polycrystals}},
volume = {38},
year = {1990}
}

@article{Liu1990,
	title = {Novel acoustic excitations in suspensions of hard-sphere colloids},
	volume = {65},
	doi = {10.1103/PhysRevLett.65.2602},
	number = {20},
	journal = {Physical Review Letters},
	publisher = {American Physical Society},
	author = {Liu, J. and Ye, L. and Weitz, D. A. and Sheng, Ping},
	year = {1990},
	pages = {2602--2605}
}

@article{Jing1991,
	title = {Theory of acoustic excitations in colloidal suspensions},
	volume = {66},
	doi = {10.1103/PhysRevLett.66.1240},
	number = {9},
	journal = {Physical Review Letters},
	author = {Jing, Xiaodun and Sheng, Ping and Zhou, Minyao},
	year = {1991},
	pages = {1240--1243}
}

@article{Alleyne1991,
	title = {A two-dimensional {Fourier} transform method for the measurement of propagating multimode signals},
	volume = {89},
	issn = {0001-4966},
	doi = {10.1121/1.400530},
	number = {3},
	journal = {The Journal of the Acoustical Society of America},
	author = {Alleyne, D. and Cawley, P.},
	month = mar,
	year = {1991},
	pages = {1159--1168},
}

@article{Page1996,
	title = {Group velocity in strongly scattering media},
	volume = {271},
	doi = {10.1126/science.271.5249.634},
	number = {5249},
	journal = {Science},
	publisher = {American Association for the Advancement of Science},
	author = {Page, J. H. and Sheng, Ping and Schriemer, H. P. and Jones, I. and Jing, Xiaodun and Weitz, D. A.},
	year = {1996},
	pages = {634--637}
}

@book{Sheng2006,
  title = {Introduction to {{Wave Scattering}}, {{Localization}} and {{Mesoscopic Phenomena}}},
  author = {Sheng, Ping},
  year = 2006,
  publisher = {Springer},
  address = {Berlin, Heidelberg},
  doi = {10.1007/3-540-29156-3},
  isbn = {978-3-540-29155-8 978-3-540-29156-5}
}

@book{Akkermans2007,
address={Cambridge},
title={Mesoscopic Physics of Electrons and Photons},
publisher={Cambridge University Press},
author={Akkermans, Eric and Montambaux, Gilles},
year={2007},
doi={10.1017/CBO9780511618833}
}

@article{Ranganathan2008,
author = {Ranganathan, Shivakumar I. and Ostoja-Starzewski, Martin},
doi = {10.1103/PhysRevLett.101.055504},
journal = {Physical Review Letters},
number = {5},
pages = {055504},
pmid = {18764407},
title = {{Universal elastic anisotropy index}},
volume = {101},
year = {2008}
}

@article{Cowan2011,
	title = {Ultrasonic wave transport in a system of disordered resonant scatterers: {Propagating} resonant modes and hybridization gaps},
	volume = {84},
	doi = {10.1103/PhysRevB.84.094305},
	number = {9},
	journal = {Physical Review B},
	author = {Cowan, M. L. and Page, J. H. and Sheng, Ping},
	year = {2011},
    pages = {094305}
}

@article{Calvet2012,
author = {Calvet, Marie and Margerin, Ludovic},
doi = {10.1121/1.3682048},
journal = {The Journal of the Acoustical Society of America},
number = {3},
pages = {1843--1862},
pmid = {22423683},
title = {{Velocity and attenuation of scalar and elastic waves in random media: A spectral function approach}},
volume = {131},
year = {2012}
}

@article{Huthwaite2014,
author = {Huthwaite, Peter},
doi = {10.1016/j.jcp.2013.10.017},
journal = {Journal of Computational Physics},
pages = {687--707},
publisher = {Elsevier Inc.},
title = {{Accelerated finite element elastodynamic simulations using the GPU}},
volume = {257},
year = {2014}
}

@article{Rokhlin2015,
author = {Rokhlin, Stanislav I. and Li, Jia and Sha, Gaofeng},
doi = {10.1121/1.4919333},
journal = {The Journal of the Acoustical Society of America},
number = {5},
pages = {2655--2669},
pmid = {25994697},
title = {{Far-field scattering model for wave propagation in random media}},
volume = {137},
year = {2015}
}

@article{Gower2019a,
	title = {Multiple waves propagate in random particulate materials},
	volume = {79},
	issn = {0036-1399},
	doi = {10.1137/18M122306X},
	number = {6},
	journal = {SIAM Journal on Applied Mathematics},
	publisher = {Society for Industrial and Applied Mathematics},
	author = {Gower, Artur L. and Parnell, William J. and Abrahams, I. David},
	year = {2019},
	pages = {2569--2592}
}

@article{Gower2019b,
	title = {A proof that multiple waves propagate in ensemble-averaged particulate materials},
	volume = {475},
	issn = {1364-5021},
	doi = {10.1098/rspa.2019.0344},
	number = {2229},
	journal = {Proceedings of the Royal Society A},
	author = {Gower, Artur L. and Abrahams, I. David and Parnell, William J.},
	year = {2019},
	pages = {20190344}
}

@article{Sha2020,
  title = {Attenuation and Velocity of Elastic Waves in Polycrystals with Generally Anisotropic Grains: Analytic and Numerical Modeling},
  author = {Sha, Gaofeng and Huang, Ming and Lowe, Michael J. S. and Rokhlin, Stanislav I.},
  year = 2020,
  journal = {The Journal of the Acoustical Society of America},
  volume = {147},
  number = {4},
  pages = {2442--2465},
  doi = {10.1121/10.0001087}
}

@article{Huang2020a,
  title = {Maximizing the Accuracy of Finite Element Simulation of Elastic Wave Propagation in Polycrystals},
  author = {Huang, Ming and Sha, Gaofeng and Huthwaite, Peter and Rokhlin, Stanislav I. and Lowe, Michael J. S.},
  year = 2020,
  journal = {The Journal of the Acoustical Society of America},
  volume = {148},
  number = {4},
  pages = {1890--1910},
  issn = {0001-4966},
  doi = {10.1121/10.0002102}
}

@phdthesis{Huang2021,
  title = {Elastic Wave Propagation in Polycrystals},
  author = {Huang, Ming},
  year = 2021,
  address = {London},
  school = {Imperial College London}
}

@article{Huang2022a,
  title = {Finite-Element and Semi-Analytical Study of Elastic Wave Propagation in Strongly Scattering Polycrystals},
  author = {Huang, Ming and Huthwaite, Peter and Rokhlin, Stanislav I. and Lowe, Michael J. S.},
  year = 2022,
  journal = {Proceedings of the Royal Society A},
  volume = {478},
  number = {2258},
  pages = {20210850},
  issn = {1364-5021},
  doi = {10.1098/rspa.2021.0850}
}

@article{Huang2022b,
  title = {Appraising Scattering Theories for Polycrystals of Any Symmetry Using Finite Elements},
  author = {Huang, Ming and Rokhlin, Stanislav I. and Lowe, Michael J. S.},
  year = 2022,
  month = sep,
  journal = {Philosophical Transactions of the Royal Society A},
  volume = {380},
  number = {2231},
  pages = {20210382},
  issn = {1364-503X},
  doi = {10.1098/rsta.2021.0382}
}

@article{Willis2024,
	title = {Energy transmission through and reflection from a layer of random composite},
	volume = {189},
	issn = {0022-5096},
	doi = {10.1016/j.jmps.2024.105689},
	journal = {Journal of the Mechanics and Physics of Solids},
	author = {Willis, J. R.},
	year = {2024},
	pages = {105689}
}

@article{Bourret1962,
  author  = {Bourret, R. C.},
  title   = {Stochastically perturbed fields, with applications to wave propagation in random media},
  journal = {Il Nuovo Cimento},
  volume  = {26},
  number  = {1},
  pages   = {1--31},
  year    = {1962},
  doi     = {10.1007/BF02754339}
}

@article{Jing1992,
  author  = {Jing, Xiaodun and Sheng, Ping and Zhou, Minyao},
  title   = {Acoustic and electromagnetic quasimodes in dispersed random media},
  journal = {Physical Review A},
  volume  = {46},
  number  = {10},
  pages   = {6513--6534},
  year    = {1992},
  doi     = {10.1103/PhysRevA.46.6513}
}

@article{Turner1999,
  author  = {Turner, Joseph A.},
  title   = {Elastic wave propagation and scattering in heterogeneous, anisotropic media: Textured polycrystalline materials},
  journal = {The Journal of the Acoustical Society of America},
  volume  = {106},
  number  = {2},
  pages   = {541--552},
  year    = {1999},
  doi     = {10.1121/1.427024}
}

@article{WeaverPao1981,
  author  = {Weaver, Richard L. and Pao, Yih-Hsing},
  title   = {Dispersion relations for linear wave propagation in homogeneous and inhomogeneous media},
  journal = {Journal of Mathematical Physics},
  volume  = {22},
  number  = {9},
  pages   = {1909--1918},
  year    = {1981},
  doi     = {10.1063/1.525164}
}

@article{ODonnell1981,
  author  = {O'Donnell, M. and Jaynes, E. T. and Miller, J. G.},
  title   = {{Kramers--Kronig} relationship between ultrasonic attenuation and phase velocity},
  journal = {The Journal of the Acoustical Society of America},
  volume  = {69},
  number  = {3},
  pages   = {696--701},
  year    = {1981},
  doi     = {10.1121/1.385566}
}

@article{Beltzer1986,
  author  = {Beltzer, Abraham I.},
  title   = {Causality and the {Keller} approximation for acoustic waves},
  journal = {Journal of Applied Physics},
  volume  = {59},
  number  = {4},
  pages   = {1456--1457},
  year    = {1986},
  doi     = {10.1063/1.336499}
}

@article{Papadakis1965,
  author  = {Papadakis, Emmanuel P.},
  title   = {Ultrasonic attenuation caused by scattering in polycrystalline metals},
  journal = {The Journal of the Acoustical Society of America},
  volume  = {37},
  number  = {4},
  pages   = {711--717},
  year    = {1965},
  doi     = {10.1121/1.1909401}
}

@article{Hirsekorn1982,
  author  = {Hirsekorn, Sigrun},
  title   = {The scattering of ultrasonic waves by polycrystals},
  journal = {The Journal of the Acoustical Society of America},
  volume  = {72},
  number  = {3},
  pages   = {1021--1031},
  year    = {1982},
  doi     = {10.1121/1.388233}
}

@article{Thompson2008,
  author  = {Thompson, R. Bruce and Margetan, F. J. and Haldipur, P. and Yu, L. and Li, A. and Panetta, P. and Wasan, H.},
  title   = {Scattering of elastic waves in simple and complex polycrystals},
  journal = {Wave Motion},
  volume  = {45},
  number  = {5},
  pages   = {655--674},
  year    = {2008},
  doi     = {10.1016/j.wavemoti.2007.09.008}
}

@article{Kube2015,
  author  = {Kube, Christopher M. and Turner, Joseph A.},
  title   = {Acoustic attenuation coefficients for polycrystalline materials containing crystallites of any symmetry class},
  journal = {The Journal of the Acoustical Society of America},
  volume  = {137},
  number  = {6},
  pages   = {EL476--EL482},
  year    = {2015},
  doi     = {10.1121/1.4921676}
}

@article{VanPamel2015,
  author  = {Van Pamel, Anton and Brett, Colin R. and Huthwaite, Peter and Lowe, Michael J. S.},
  title   = {Finite element modelling of elastic wave scattering within a polycrystalline material in two and three dimensions},
  journal = {The Journal of the Acoustical Society of America},
  volume  = {138},
  number  = {4},
  pages   = {2326--2336},
  year    = {2015},
  doi     = {10.1121/1.4931445}
}

@article{VanPamel2017,
  author  = {Van Pamel, Anton and Sha, Gaofeng and Rokhlin, Stanislav I. and Lowe, Michael J. S.},
  title   = {Finite-element modelling of elastic wave propagation and scattering within heterogeneous media},
  journal = {Proceedings of the Royal Society A},
  volume  = {473},
  number  = {2197},
  pages   = {20160738},
  year    = {2017},
  doi     = {10.1098/rspa.2016.0738}
}

@article{Ryzy2018,
  author  = {Ryzy, Martin and Grabec, Tom{\'a}{\v{s}} and Sedl{\'a}k, Petr and Veres, Istv{\'a}n A.},
  title   = {Influence of grain morphology on ultrasonic wave attenuation in polycrystalline media with statistically equiaxed grains},
  journal = {The Journal of the Acoustical Society of America},
  volume  = {143},
  number  = {1},
  pages   = {219--229},
  year    = {2018},
  doi     = {10.1121/1.5020785}
}

@article{Quey2011,
  author  = {Quey, R. and Dawson, P. R. and Barbe, F.},
  title   = {Large-scale {3D} random polycrystals for the finite element method: {Generation}, meshing and remeshing},
  journal = {Computer Methods in Applied Mechanics and Engineering},
  volume  = {200},
  number  = {17--20},
  pages   = {1729--1745},
  year    = {2011},
  doi     = {10.1016/j.cma.2011.01.002}
}

@book{Sato2012,
  author    = {Sato, Haruo and Fehler, Michael C. and Maeda, Takuto},
  title     = {Seismic Wave Propagation and Scattering in the Heterogeneous Earth},
  edition   = {2nd},
  publisher = {Springer},
  address   = {Berlin, Heidelberg},
  year      = {2012},
  doi       = {10.1007/978-3-642-23029-5}
}

@article{Hoffman2012,
  author  = {Hoffman, Joseph J. and Nelson, Amber M. and Holland, Mark R. and Miller, James G.},
  title   = {Cancellous bone fast and slow waves obtained with {Bayesian} probability theory correlate with porosity from computed tomography},
  journal = {The Journal of the Acoustical Society of America},
  volume  = {132},
  number  = {3},
  pages   = {1830--1837},
  year    = {2012},
  doi     = {10.1121/1.4739455}
}

@article{Meziere2016,
  author  = {M{\'e}zi{\`e}re, Fabien and Juskova, Petra and Woittequand, Jonathan and Muller, Marie and Bossy, Emmanuel and Boistel, Renaud and Malaquin, Laurent and Derode, Arnaud},
  title   = {Experimental observation of ultrasound fast and slow waves through three-dimensional printed trabecular bone phantoms},
  journal = {The Journal of the Acoustical Society of America},
  volume  = {139},
  number  = {2},
  pages   = {EL13--EL18},
  year    = {2016},
  doi     = {10.1121/1.4939297}
}

@article{Gower2021,
  author  = {Gower, Artur L. and Kristensson, Gerhard},
  title   = {Effective waves for random three-dimensional particulate materials},
  journal = {New Journal of Physics},
  volume  = {23},
  pages   = {063083},
  year    = {2021},
  doi     = {10.1088/1367-2630/abdfee}
}

@unpublished{Huang2026prl,
  author  = {Huang, Ming and Rokhlin, Stanislav I. and Lowe, Michael J. S.},
  title   = {Multiple propagating modes in random polycrystals},
  note    = {under review in Physical Review Letters},
  year    = {2026},
  doi     = {10.48550/arXiv.2610.24047},
  eprint  = {2610.24047},
  archivePrefix = {arXiv}
}

@unpublished{Huang2026jasa,
  author = {Huang, Ming},
  title  = {Multimode dispersion, attenuation and amplitude from line-shape fitting in the frequency--wavenumber plane},
  note   = {{Zenodo} preprint, doi:10.5281/zenodo.23493994, under review in The Journal of the Acoustical Society of America},
  year   = {2026},
  doi     = {10.5281/zenodo.23493994},
  url     = {https://doi.org/10.5281/zenodo.23493994}
}

@article{Roy2025,
  author  = {Roy, Anubhav and Kube, Christopher M.},
  title   = {Multiple scattering of elastic waves in polycrystals},
  journal = {Journal of the Mechanics and Physics of Solids},
  volume  = {203},
  pages   = {106237},
  year    = {2025},
  doi     = {10.1016/j.jmps.2025.106237}
}

@article{Lan2015a,
  author  = {Lan, Bo and Lowe, Michael J. S. and Dunne, Fionn P. E.},
  title   = {A spherical harmonic approach for the determination of {HCP} texture from ultrasound: {A} solution to the inverse problem},
  journal = {Journal of the Mechanics and Physics of Solids},
  volume  = {83},
  pages   = {179--198},
  year    = {2015},
  doi     = {10.1016/j.jmps.2015.06.014}
}

@article{Lan2015b,
  author  = {Lan, Bo and Lowe, Michael J. S. and Dunne, Fionn P. E.},
  title   = {A generalized spherical harmonic deconvolution to obtain texture of cubic materials from ultrasonic wave speed},
  journal = {Journal of the Mechanics and Physics of Solids},
  volume  = {83},
  pages   = {221--242},
  year    = {2015},
  doi     = {10.1016/j.jmps.2015.06.012}
}

@article{Rokhlin2021,
  author  = {Rokhlin, Stanislav I. and Sha, Gaofeng and Li, Jia and Pilchak, Adam L.},
  title   = {Inversion methodology for ultrasonic characterization of polycrystals with clusters of preferentially oriented grains},
  journal = {Ultrasonics},
  volume  = {115},
  pages   = {106433},
  year    = {2021},
  doi     = {10.1016/j.ultras.2021.106433}
}

@article{Calvet2008,
  author  = {Calvet, Marie and Margerin, Ludovic},
  title   = {Constraints on grain size and stable iron phases in the uppermost inner core from multiple scattering modeling of seismic velocity and attenuation},
  journal = {Earth and Planetary Science Letters},
  volume  = {267},
  number  = {1-2},
  pages   = {200--212},
  year    = {2008},
  doi     = {10.1016/j.epsl.2007.11.048}
}

@article{Monnereau2010,
  author  = {Monnereau, Marc and Calvet, Marie and Margerin, Ludovic and Souriau, Annie},
  title   = {Lopsided growth of {E}arth's inner core},
  journal = {Science},
  volume  = {328},
  number  = {5981},
  pages   = {1014--1017},
  year    = {2010},
  doi     = {10.1126/science.1186212}
}

@article{Calvet2016,
  author  = {Calvet, Marie and Margerin, Ludovic},
  title   = {Impact of grain shape on seismic attenuation and phase velocity in cubic polycrystalline materials},
  journal = {Wave Motion},
  volume  = {65},
  pages   = {29--43},
  year    = {2016},
  doi     = {10.1016/j.wavemoti.2016.04.001}
}

@article{Pang2023,
  author  = {Pang, Guanning and Koper, Keith D. and Wu, Sin-Mei and Wang, Wei and Lasbleis, Marine and Euler, Garrett},
  title   = {Enhanced inner core fine-scale heterogeneity towards {E}arth's centre},
  journal = {Nature},
  volume  = {620},
  pages   = {570--575},
  year    = {2023},
  doi     = {10.1038/s41586-023-06213-2}
}

@article{Rokhlin1992,
  author  = {Rokhlin, S. I. and Bolland, T. K. and Adler, Laszlo},
  title   = {High-frequency ultrasonic wave propagation in polycrystalline materials},
  journal = {The Journal of the Acoustical Society of America},
  volume  = {91},
  number  = {1},
  pages   = {151--165},
  year    = {1992},
  doi     = {10.1121/1.402764}
}

@article{Tatarskii1963,
  author  = {Tatarskii, V. I. and Gertsenshtein, M. E.},
  title   = {Propagation of waves in a medium with strong fluctuation of the refractive index},
  journal = {Soviet Physics JETP},
  volume  = {17},
  number  = {2},
  pages   = {458},
  year    = {1963},
  url     = {https://jetp.ras.ru/cgi-bin/e/index/e/17/2/p458?a=list},
  note    = {[Zh. Eksp. Teor. Fiz. 44, 676 (1963)]}
}

@article{Kube2015sc,
  author  = {Kube, Christopher M. and Turner, Joseph A.},
  title   = {Ultrasonic attenuation in polycrystals using a self-consistent approach},
  journal = {Wave Motion},
  volume  = {57},
  pages   = {182--193},
  year    = {2015},
  doi     = {10.1016/j.wavemoti.2015.04.002}
}

@article{Kube2016,
  author  = {Kube, Christopher M.},
  title   = {Elastic anisotropy of crystals},
  journal = {AIP Advances},
  volume  = {6},
  pages   = {095209},
  year    = {2016},
  doi     = {10.1063/1.4962996}
}

@article{KubeDeJong2016,
  author  = {Kube, Christopher M. and de Jong, Maarten},
  title   = {Elastic constants of polycrystals with generally anisotropic crystals},
  journal = {Journal of Applied Physics},
  volume  = {120},
  pages   = {165105},
  year    = {2016},
  doi     = {10.1063/1.4965867}
}

@book{Auld1973,
  author    = {Auld, B. A.},
  title     = {Acoustic Fields and Waves in Solids},
  volume    = {1},
  publisher = {Wiley},
  address   = {New York},
  year      = {1973}
}

@article{Yang2012,
  author  = {Yang, L. and Lobkis, O. I. and Rokhlin, S. I.},
  title   = {An integrated model for ultrasonic wave propagation and scattering in a polycrystalline medium with elongated hexagonal grains},
  journal = {Wave Motion},
  volume  = {49},
  pages   = {544--560},
  year    = {2012},
  doi     = {10.1016/j.wavemoti.2012.03.003}
}

\end{document}